\documentclass[sigconf]{acmart}
\AtBeginDocument{%
  }

\usepackage{arydshln}
\usepackage{textcomp}
\usepackage{soul, color, xcolor}
\usepackage{float}
\usepackage[utf8]{inputenc}
\usepackage{array}

\newcommand{\takeaway}[1]{
\noindent\rule{\linewidth}{0.1pt}
\par\nobreak\noindent\textbf{Summary of Results:}
#1
\vspace{-2mm}
\par\nobreak\noindent
\rule{\linewidth}{0.1pt}
}

\newcommand{\lowTrustHighDistrust}{low-trust-high-distrust}
\newcommand{\lowTrustLowDistrust}{low-trust-low-distrust}
\newcommand{\highTrustHighDistrust}{high-trust-high-distrust}
\newcommand{\highTrustLowDistrust}{high-trust-low-distrust}

\def\mybar#1{
  {\color{gray}\rule{#1cm}{6pt}}} 

\makeatletter
\def\adl@drawiv#1#2#3{%
        \hskip.5\tabcolsep
        \xleaders#3{#2.5\@tempdimb #1{1}#2.5\@tempdimb}%
                 #2\z@ plus1fil minus1fil\relax 
        \hskip.5\tabcolsep}
\newcommand{\cdashlinelr}[1]{%
  \noalign{\vskip\aboverulesep
          \global\let\@dashdrawstore\adl@draw
          \global\let\adl@draw\adl@drawiv}
  \cdashline{#1}
  \noalign{\global\let\adl@draw\@dashdrawstore
          \vskip\belowrulesep}}
\makeatother

\begin{document}

\title[]{Understanding the Practices, Perceptions, and (Dis)Trust of Generative AI among Instructors: A Mixed-methods Study in the U.S. Higher Education}


\author{Wenhan Lyu}
\affiliation{
    \institution{William \& Mary}
    \city{Williamsburg}
    \state{VA}
    \country{USA}} 
\email{wlyu@wm.edu}
\orcid{0009-0004-9129-8689} 

\author{Shuang Zhang}
\authornote{This work was completed while serving as a research assistant intern at Human-Computer Interaction Lab, William \& Mary.}
\affiliation{
    \institution{Renmin University of China}
    \city{Beijing}
    \country{China}} 
\email{zshuang2000@ruc.edu.cn}
\orcid{0000-0003-1618-2163}

\author{Tingting (Rachel) Chung}  
\affiliation{
    \institution{William \& Mary}
    \city{Williamsburg}
    \state{VA}
    \country{USA}}
\email{rachel.chung@mason.wm.edu}
\orcid{0000-0002-0250-4873}

\author{Yifan Sun}  
\affiliation{
    \institution{William \& Mary}
    \city{Williamsburg}
    \state{VA}
    \country{USA}}
\email{ysun25@wm.edu}
\orcid{0000-0003-3532-6521}

\author{Yixuan Zhang}  
 \affiliation{
    \institution{William \& Mary}
    \city{Williamsburg}
    \state{VA}
    \country{USA}} 
\email{yzhang104@wm.edu}
\orcid{0000-0002-7412-4669}

\renewcommand{\shortauthors}{Lyu et al.}

\begin{abstract} 
Generative AI (GenAI) has brought opportunities and challenges for higher education as it integrates into teaching and learning environments. As instructors navigate this new landscape, understanding their engagement with and attitudes toward GenAI is crucial. We surveyed 178 instructors from a single U.S. university to examine their current practices, perceptions, trust, and distrust of GenAI in higher education in March 2024. While most surveyed instructors reported moderate to high familiarity with GenAI-related concepts, their actual use of GenAI tools for direct instructional tasks remained limited. Our quantitative results show that trust and distrust in GenAI are related yet distinct; high trust does not necessarily imply low distrust, and vice versa. We also found significant differences in surveyed instructors' familiarity with GenAI across different trust and distrust groups. Our qualitative results show nuanced manifestations of trust and distrust among surveyed instructors and various approaches to support calibrated trust in GenAI. We discuss practical implications focused on (dis)trust calibration among instructors.
\end{abstract}



\maketitle

\section{Introduction}  

Generative AI (GenAI) is rapidly transforming teaching and learning in higher education, introducing significant changes and uncertainties~\cite{michel2023challenges}. Faculty, students, and institutions all face uncertainty and anxiety surrounding the role of GenAI in teaching and learning, as it is uncertain to what extent they should embrace or restrict the use of GenAI in educational contexts~\cite{ adeshola2023opportunities}. Policymakers also face challenges in devising appropriate regulatory frameworks and guidelines to manage the integration of GenAI into higher education, balancing innovation with ethical considerations and academic integrity~\cite{luo2024critical}. Meanwhile, attitudes and practices among different stakeholders regarding GenAI vary significantly across educational communities. For example, according to a survey conducted in 2023 by Tyton Partners, most students are increasingly curious about GenAI, with nearly half of the college students using these tools regularly~\cite{tyton_press_2023}. In contrast, only 22\% of faculty members have adopted GenAI~\cite{Coffey_2023}. Even among instructors, the integration of GenAI into educational practices has sparked polarized reactions~\cite{mishra2024generative, dagostino_2023, polarizedAI2023}. Such a split often manifests as a division between those who view traditional educational methods as outdated---labeling skeptics and distrust as ``\emph{you're a dinosaur}~\cite{polarizedAI2023}''---and those who caution against an unrestrained embrace of AI, fearing it could undermine fundamental educational values, such as critical thinking and academic integrity~\cite{van2023chatgpt, sullivan2023chatgpt}. These varying practices and attitudes shed light on a broader discourse on educators' perceptions, attitudes, and trust towards GenAI, as these perspectives could help explain their practices and approaches to GenAI use and integration in teaching and learning. 

An emerging body of work has explored current practices of using GenAI in educational contexts to explore the use cases of embracing the potential of GenAI and identify the risks and challenges involved in GenAI, turning to both students'~\cite{amoozadeh2024trust} and instructors'~\cite{ghimire2024generative} perceptions and attitudes to GenAI in higher education. However, fewer studies have focused on unpacking the dynamics of trust and distrust among instructors. 
Trust in GenAI, in our study context, can be viewed as an individual's willingness to rely on the system and accept the accompanying vulnerabilities, grounded in the belief that GenAI can reliably enhance educational outcomes (e.g., providing accurate insights, personalizing learning). Distrust, on the other hand, arises when GenAI is perceived as unreliable, harmful, or misaligned with instructional goals, causing educators or learners to withhold reliance or limit usage. And yet, instructors' trust and distrust can significantly influence their decisions to adopt GenAI, how they use it in their teaching, and their overall attitudes and practices, which in turn shape students' perceptions and trust in GenAI~\cite{luo2024critical, donnell2009relationship, hall2013assessing, palmore2011faculty, bunk2015understanding, kosak2004prepared}. 

Understanding the factors and considerations contributing to the formation of (dis)trust among educators will also provide insights into how to cultivate a balanced perspective, preventing and avoiding \emph{blind trust} (i.e., characterized by an uncritical acceptance of GenAI's capabilities without proper assessment of potential risks) and \emph{blind distrust}, which involves a complete rejection of GenAI's potential benefits even without explicit reasons, direct experience, or understanding of GenAI. Such extremes can lead to a trust crisis, a situation where a significant loss of trust or confidence among stakeholders occurs, affecting the overall educational environment and its ability to embrace technological innovations~\cite{selwyn2013distrusting}. Additionally, the conceptual clarity of trust and distrust in GenAI remains ambiguous. While previous research often views trust and distrust as two extremes of a single dimension or uses these two terms interchangeably---implying high trust \textit{means} low distrust---some scholars have argued that trust and distrust can be two independent constructs in certain contexts~\cite{lankton2015technology, zhang2024profiling}. In other words, people can hold trust and distrust at the same time, influenced by different sets of factors and considerations. And thus, we believe understanding distrust in GenAI deserves as much attention as trust, within the context of education, where both could influence technology adoption and use practices. 

This work makes the following contributions:
\begin{enumerate}
    \item We conducted a mixed-methods study with 178 instructors from a mid-Atlantic U.S. university, examining their current practices of, perceptions of, and trust/distrust in GenAI, and providing new empirical insights within a culturally specific context.
    \item We analyzed the coexisting manifestations of trust and distrust in GenAI from surveyed instructors, demonstrating that trust and distrust can be related yet distinct.
    \item We proposed design implications for fostering calibrated (dis)trust in GenAI for educational purposes, focusing on practical strategies to support instructors' informed and balanced engagement with this technology.
\end{enumerate} 

\section{Related Work} 
\subsection{Instructors' Perceptions and Practices of Technology}
Since the concept of using computers and other information and communication technologies in education emerged~\cite{nwana1990intelligent}, researching technology acceptance in teaching and learning contexts has become an attractive and consistent trend~\cite{imtiaz2014review, teo2011technology}. Prior research has investigated acceptance and attitudes toward learning technology in educational contexts, with a primary focus on university students rather than instructors~\cite{granic2019technology}. However, it is instructors' attitudes that can significantly impact the acceptance of new technologies when incorporated into university education~\cite{donnell2009relationship, hall2013assessing, palmore2011faculty, bunk2015understanding, kosak2004prepared}. 
Compared to a generally positive perception by instructors of technology~\cite{marzilli2014faculty, kopcha2016understanding, qudais2010senior, onwuagboke2016faculty, akbarilakeh2019attitudes}, a decline in acceptance of technology is often observed among faculty when sufficient support from institutions are not provided~\cite{ramlo2021coronavirus, daumiller2021shifting}, demonstrating and emphasizing the importance of understanding faculty attitudes toward technology early to formulate related policies or provide proper training by institutions.

The release of ChatGPT in 2022 brought GenAI-based applications into the spotlight and attracted widespread attention. GenAI continues driving innovation in higher education, presenting both opportunities and challenges for teaching and learning~\cite{michel2023challenges, adeshola2023opportunities}. On the one hand, a growing body of literature has documented the potential of integrating GenAI into higher education, such as generating course materials~\cite{denny2022robosourcing, sarsa2022automatic} and assisting students' programming tasks~\cite{lyu2024evaluating, poldrack2023ai}. On the other hand, recent work has also highlighted issues of the use of GenAI in higher education~\cite{kasneci2023chatgpt}, such as accuracy~\cite{johnson2023assessing}, reliability~\cite{johnson2023assessing, walker2023reliability}, and plagiarism~\cite{jarrah2023using}. 

With the rapid adoption of GenAI across different fields, recent studies have looked into people's attitudes and practices regarding using GenAI, particularly in the context of higher education.  For example, current research has explored both undergraduate and graduate students' perceptions of ChatGPT across cultural contexts, including Asia~\cite{ngo2023perception, farhi2023analyzing, shoufan2023exploring}, Europe~\cite{stohr2024perceptions, singh2023exploring, romero2023use}, Australia ~\cite{gruenhagen2024rapid}, and North America~\cite{baek2024chatgpt}. These investigations identify shared ethical concerns regarding GenAI while also revealing region-specific attitudes, suggesting the importance of culturally contextualized approaches to its integration. Likewise, Luo~\cite{luo2024critical} examined GenAI-related institutional policies across universities worldwide, revealing significant inconsistencies and pointing to the need for more cohesive, context-aware frameworks to guide GenAI adoption in educational settings. 

However, fewer studies have explicitly examined instructors' perceptions and practices of GenAI~\cite{albayati2024investigating}. A few exceptional examples include recent survey studies conducted in Europe~\cite{beege2024ai,kiryakova2023chatgpt}, Asia~\cite{espartinez2024exploring}, North America~\cite{amani2023generative, ghimire2024generative}, and mixed cultural contexts~\cite{lau2023ban}. Among this body of work, they mostly focus on perceived benefits (e.g., enhanced teaching efficiency) and risks (e.g., unethical usage, bias, privacy concerns) rather than systematically differentiating trust from distrust. For example, a survey of German STEM teachers by Beege et al.~\cite{beege2024ai} revealed that while perceived competence in ChatGPT supported usage intentions, perceived risks and concerns negatively impacted adoption, suggesting the nuanced interplay between confidence and hesitation. Similarly, Bulgarian professors exhibited predominantly positive attitudes toward ChatGPT but nonetheless expressed concerns about unethical applications~\cite{kiryakova2023chatgpt}. Meanwhile, Espartinez~\cite{espartinez2024exploring} used Q-methodology in the Philippines to highlight a diverse spectrum of instructor and student perspectives, ranging from ethical considerations to practical challenges. North American studies also contribute to this discussion, where Amani et al.~\cite{amani2023generative}, for example, surveyed faculty and staff on overall perceptions of GenAI but did not explicitly dissect trust versus distrust. Furthermore, much of this work remains confined to single disciplines~\cite{ayanwale2024exploring, lau2023ban}, leaving a gap in understanding how instructors across a broad range of fields relate to GenAI. Extending the existing body of literature, our work seeks to unpack how trust and distrust in GenAI manifest among instructors in higher education.

\subsection{Trust and Distrust in AI-Driven Educational Environments}

Trust is often conceptualized as a willingness to be vulnerable based on positive expectations of another party’s intentions or behavior, enabling cooperation and reducing uncertainty~\cite{mayer1995integrative, dirks2001role}. In contrast, distrust is conceptualized as a distinct construct reflecting suspicion and the expectation of harm~\cite{lewicki1998trust}. While trust is generally viewed positively as a facilitator of relationships and systems, distrust is often associated with caution and avoidance. Some scholars argue that trust and distrust can coexist, highlighting the nuanced and dynamic ways individuals evaluate others or systems~\cite{lewicki1998trust, schilke2021trust}. These conceptualizations of trust and distrust have been extended to interactions with AI systems. For example, Jacovi et al.~\cite{jacovi2021formalizing} define trust and distrust in AI as a human perceiving an AI model as trustworthy and accepting vulnerability to the model's actions or not. Prior research has explored different factors contributing to individuals' trust in GenAI, such as the AI models' trustworthiness~\cite{sun2024trustllm}, explainability~\cite{bhattacharjee2024towards}, transparency~\cite{sarker2024llm}, etc. However, research has also shown that trust is highly contextual, with users basing their trust on different factors depending on the task. For example, preciseness plays a key role in shaping trust in AI recommendations~\cite{kim2021you}, whereas users prioritize the originality of AI-generated content in creative tasks~\cite{daly2024sensemaking}. Furthermore, individual differences in trust have also been well-documented since the pioneering research on technology acceptance~\cite{davis1989perceived, davis1989technology}. Thus, it is crucial to investigate how trust and distrust factors influence the adoption of technologies like GenAI within specific groups, such as instructors, whose professional roles and dedicated tasks may shape their engagement with these technologies.

A body of research has already focused on analyzing and explaining trust in GenAI within the context of higher education. For example, Amoozadeh et al.~\cite{amoozadeh2024trust} conducted an exploratory study explicating university students' trust in GenAI, while Kim et al.~\cite{kim2023you} found that subpar responses from GenAI significantly decreased users' trust. However, understanding distrust is just as important as trust, as both factors influence how instructors engage with and integrate technologies like GenAI. Although much of the existing research prioritizes trust, and some recent studies have started to explore distrust~\cite{lankton2015technology, zhang2024profiling}, the role of distrust in the adoption of GenAI remains underexplored and insufficiently acknowledged. Given the imperfections of GenAI models, fostering a healthy level of distrust is essential for responsible use, underscoring the need to consider both trust and distrust as equally important factors~\cite{peters2023importance}. Meanwhile, it is essential to avoid blind trust, which involves unconditionally accepting without critical evaluation, as this can leave individuals vulnerable to exploitation in such situations~\cite{min2023development}. Conversely, blind distrust occurs when trust is withheld despite evidence of trustworthiness, potentially leading individuals to miss valuable opportunities~\cite{beccerra1999trust}. 

Although our discussion centers on GenAI, we situate it within the broader context of technology-driven educational tools. Traditional views of trust in education often tend to categorize trust in educational settings into inter-organizational trust and interpersonal trust~\cite{niedlich2021comprehensive}, the rise of Intelligent Tutoring Systems (ITS)~\cite{nwana1990intelligent, sleeman1982intelligent} has introduced a new dimension of trust between humans and educational tools in modern educational environments. Existing research on teachers' trust in GenAI-based educational tools largely centers on K-12 education~\cite{nazaretsky2022teachers, viberg2024explains, qin2020understanding, nazaretsky2022instrument, beege2024ai}. In higher education, Wang et al.~\cite{wang2020participant} examined Chinese instructors’ attitudes toward ITS, though their work emphasized identifying relevant factors rather than examining how trust dynamics are formed. Similarly, Klein et al.~\cite{klein2019technological} studied instructors’ engagement with learning tools in a context culturally similar to ours, but conducted their study before the emergence of GenAI, leaving AI-related issues unexamined. Our work seeks to address this research gap by examining the nuanced ways trust and distrust can manifest among instructors using GenAI, and by offering actionable insights for (dis)trust calibration among both instructors and students.

This work aims to answer the following research questions (RQs): \\
    \textbf{RQ1.} What are the current practices of instructors in the U.S. higher education using (and not using) GenAI in the higher educational context? and \\
    \textbf{RQ2.} How do instructors in the U.S. higher education trust and distrust GenAI in an educational context? and how trust and distrust in GenAI manifest among them?

\section{Method}   
Upon approval from our institution's Institutional Review Board (IRB), we conducted a survey study with 178 instructors at a research-focused 4-year university located in the Mid-Atlantic region of the United States in March 2024. The university offers a diverse range of undergraduate and graduate programs across disciplines such as arts, sciences, business, and law, and is known for its emphasis on liberal arts education. While the institution has policies and practices related to teaching and technology integration, there were no formal guidelines specifically addressing the use of GenAI at the time of the study. Below, we describe the survey recruitment approach, participants overview, survey design and measurements, and data analysis methods. 

\subsection{Survey Study Recruitment \& Participants Overview}
The research team provided the survey link to the provost's office, and then the provost's office sent out the survey to institutional mailing lists, ensuring that instructors from all academic departments across the university were invited to participate. The recruitment emails outlined the purpose of the study and a link to the survey. The survey was open for four weeks, during which reminders were sent twice to maximize participation rates. In total, we reached out to approximately 1000 faculty members, from which we received 219 responses. After removing incomplete responses, we left 178 responses for data analysis. \autoref{tab:demographic} shows an overview of the demographic information of our participants. 

\begin{table*}
\centering
  \caption{Demographic characteristics of our participants.}
  \label{tab:demographic}
  \footnotesize
  \begin{tabular}{llrrl} 
    \toprule
        \textbf{Demographic} & \textbf{Response Options} & \textbf{Number of Participants} & \textbf{Percentage} & \\ 
        & &  (Total N = 178) & \% & \\ 
    \midrule
    Gender & Man         & 93  & 52.2\%  & \mybar{.522} \\  
           & Woman	      & 68  & 38.2\%  & \mybar{.382} \\
           & Non-binary  & 1   & 0.6\%   & \mybar{.006} \\
           & Prefer not to say	& 16   & 9\%  & \mybar{.09} \\
    \midrule
    Age & 25-34	& 16 & 9\% & \mybar{.09}  \\
         & 35-44	   & 44 & 24.7\% & \mybar{.247}  \\
         & 45-54	   & 41 & 23\% & \mybar{.23}  \\
         & 55-64	   & 37 & 20.8\% & \mybar{.208}  \\   
         & 65-74	   & 18 & 10.1\% & \mybar{.101}  \\  
         & 75+      & 2 & 1.1\% & \mybar{.011}  \\  
         & Prefer not to say  & 20 & 11.2\% & \mybar{.112}  \\ 
         \midrule
    Education   & PhD/Doctoral degree  & 142 & 79.8\% & \mybar{.798}  \\
                & Master's degree	   & 22 & 12.4\%  & \mybar{.124}  \\
                & Prefer not to say    & 14 & 7.8\%   & \mybar{.078} \\
    \midrule
    Race & African American or Black              & 5 & 2.8\% & \mybar{.028}  \\
    & American Indian or Alaskan Native, White & 1 & 0.6\% & \mybar{.006}  \\
    & Asian & 17    & 9.6\%         & \mybar{.096}  \\
    & Asian, White  & 1 & 0.6\%     & \mybar{.006}  \\
    & Multi-racial  & 6 & 3.4\%     & \mybar{.034}  \\
    & White         & 132 & 74.2\%  & \mybar{.742}  \\
    & Other         & 3 & 1.7\%     & \mybar{.017}  \\       
    \midrule
    Ethnicity   & Hispanic or Latino      &  13   & 7.3\%  & \mybar{.073} \\
    & Not Hispanic or Latino &  156   & 87.6\%  & \mybar{.876} \\
    & Prefer not to say    &  9   & 5.1\%  & \mybar{.051} \\
    \midrule
     Year of teaching & 0-10 years & 53 & 29.8\% & \mybar{.298}  \\
         & 11-20	years   & 66 & 37.1\% & \mybar{.371}  \\
         & 21-30 years   & 32 & 18\%   & \mybar{.18}  \\
         & 31-40 years   & 9  & 5.1\%  & \mybar{.051}  \\
         & 41+ years     & 4  & 2.2\%  & \mybar{.022}  \\
         & Prefer not to say            & 14 & 7.9\%  & \mybar{.079}  \\
    \midrule
    Disciplines & Arts and Sciences & 114 & 64.0\% & \mybar{.640} \\
        & School of Business & 25 & 14.0\% & \mybar{.140} \\
        & School of Education & 13 & 7.3\% & \mybar{.073} \\
        & School of Law & 12 & 6.8\% & \mybar{.068} \\
        & School of Marine Science & 8 & 4.5\% & \mybar{.045} \\
        & Anonymous & 6 & 3.4\% & \mybar{.034} \\
  \bottomrule   
\end{tabular} 
\end{table*}

\subsection{Survey Design and Measurements}
The survey includes four blocks of questions: 1) practices of and familiarity with GenAI, 2) attitudes and trust towards GenAI in teaching and learning, 3) demographic information, and 4) open-ended questions for additional insights. Detailed survey questions can be found in the supplemental materials. 

\subsubsection{Prior GenAI Experience} 
We ask several questions to understand instructors' current practices of GenAI and plans regarding it for teaching purposes, including 1) instructors' current practices and intentions of using GenAI across different instructional activities, 2) types of instruction tasks instructors typically use GenAI tools, and 3) attitudes towards GenAI training to understand instructors' confidence in their own and their students' readiness to effectively use GenAI.

\subsubsection{Attitudes, Trust, and Distrust of GenAI in Teaching and Learning}
Questions in this block focus on instructors' attitudes, levels of trust, and distrust in integrating GenAI technologies in academic environments. Building on previous research~\cite{dorton2022adaptations, choudhury2023investigating, wang2024critical, yusuf2024implementing}, we examined trust in GenAI in terms of their efficiency, effective personalize learning, adapt teaching methods to enhance educational outcomes, inspire them to incorporate GenAI into their courses, require changes in curriculum design for GenAI integration, and enhance student problem-solving skills. Conversely, we evaluated distrust by investigating whether instructors view GenAI as a threat to student mental health, a potential source of misinformation, a factor in skill degradation, a concern for the accuracy of content, a limitation on student creativity, and a deterrent to independent problem-solving~\cite{jaidka2024misinformation, abdelwahab2023business}. \autoref{tab:trust_distrust_questions_list} shows the statements and their corresponding dimensions of measurement.

\begin{table*}[h!]
\centering
\caption{Constructs, dimensions, and statements on instructors' trust and distrust in GenAI. All measurements were presented as five-level Likert scales, and participants could choose an answer from ``Strongly disagree'', ``Somewhat disagree'', ``Neutral'', ``Somewhat agree'', or ``Strongly agree'' to each statement. }
\begin{tabular}{llp{12cm}}
    \toprule
    \textbf{Construct} & \textbf{Dimension} & \textbf{Statement} \\
    \midrule
    Trust   & Competence~\cite{chan2023deconstructing}     & GenAI can make college instructors more efficient.\\
            & Personalization~\cite{khan2024path}& GenAI allows me to implement personalized learning effectively in my courses. \\
            & Adaptability~\cite{choudhuri2024guides}    & I need to change how I teach my classes because GenAI can enhance education outcomes.\\
            & Anticipation~\cite{saunders2024making}   & I am excited about the opportunity to incorporate GenAI into my courses.\\
            & Transformation~\cite{zhai2024transforming}  & There is a need to change curriculum design to integrate GenAI to enhance educational outcomes.\\
            & Enrichment~\cite{zhou2024mediating}      & Using GenAI enhances student problem-solving skills.\\
    \midrule
    Distrust    & Malevolence~\cite{olohunfunmi2024exploring} &  GenAI is a threat to student mental health in my courses. \\
                & Dishonesty~\cite{yusuf2024generative}  &  GenAI is a potential source of misinformation for students in my courses. \\
                & Skepticism~\cite{li2024user}  &  I feel that students’ own skills degrade when they use GenAI extensively. \\
                & Inaccuracy~\cite{yang2024genai}  &  I am concerned about the accuracy of content produced by GenAI. \\
                & Restriction~\cite{akbar2024revolutionizing} &  Using GenAI restricts students’ own creativity. \\
                & Demotivation~\cite{dai2024students} &  Using GenAI makes students less inclined to solve problems on their own. \\
    \bottomrule
\end{tabular}
\label{tab:trust_distrust_questions_list}
\end{table*}

In addition, instructors' familiarity with GenAI likely influences their initial levels of trust or distrust in the technology. Considering that different teaching levels require varied approaches and attitudes toward GenAI to achieve expected learning outcomes and course structures, instructors' initial trust or distrust in GenAI might vary based on the academic level they primarily teach. These two questions were thus included in our analysis: 1) \textit{How familiar are you with Generative Artificial Intelligence (GenAI)?} with response options ranging from ``Not at all familiar'' to ``Very familiar'', and 2) \textit{Do you teach primarily in undergraduate or graduate courses? Select both if you teach both}, with response options of undergraduate and graduate levels.

\subsubsection{Demographics \& Background Information}
We also asked for participants' demographic information (e.g., gender, age, race and ethnicity, education) and teaching experience (e.g., years of teaching, undergraduate/graduate level taught), while school/college affiliation was recorded from the account used to complete the survey. 

\subsubsection{Open-ended Questions}
In addition to structured survey items, we included open-ended questions to capture more nuanced perspectives on GenAI. Participants were invited to share specific positive or negative examples from their experiences using GenAI in instructional settings (e.g., classroom activities, assignments, assessments, academic integrity issues, curriculum redesign, or other relevant educational practices). Furthermore, instructors were encouraged to discuss their concerns regarding using GenAI in teaching and learning, which could help us understand the motivation behind instructors' attitudes, trust, and distrust towards GenAI.

\subsection{Data Analysis}
\label{subsec:method_data_analysis}

\textbf{Quantitative Data Analysis: }
We used several statistical analyses to investigate instructors' perceptions and attitudes toward Gen AI in teaching and learning. First, we depicted instructors' GenAI experience through descriptive statistics. Second, we employed Welch's ANOVA (using R package ggstatsplot~\cite{patil2018ggstatsplot}) to examine instructors' attitudes from the trust and distrust perspectives. The expected sample sizes in our study were calculated by G*Power ~\cite{kang2021sample} software. We used ANOVA to examine the differences in familiarity with GenAI across different groups, and the calculated total sample size was 140 (effect size=0.3, $\alpha$=0.05, power=0.8, number of groups=5). Similarly, we used ANOVA to explore the differences in trust and distrust across different teaching levels (undergraduate vs. graduate level), and the calculated total sample size was 144 (effect size=0.3, Bonferroni correction $\alpha$=0.017, power=0.8, number of groups=3). The actual number of valid responses in our study was 178, indicating an acceptable sample size. \\ 
\textbf{Qualitative Data Analysis: }
We conducted a qualitative analysis with the open-ended responses using the General Inductive Approach~\cite{thomas2006general}. The data included 294 valid responses, with a total word count of 10,269 words. According to the thematic analysis guideline, two researchers first read the answers to open-ended questions to have an initial understanding of the instructors' trust and distrust in GenAI learning and teaching; then, they created low-level codes to identify all sentences and phrases related to instructors' trust and distrust, using Nvivo~\cite{wong2008data}. Subsequently, two researchers integrated related low-level codes to achieve high-level themes.  The two researchers resolved any discrepancies or disagreements through discussions to reach a consensus on the interpretation of the data. All authors met regularly, discussed the emerging themes, and iterated the themes throughout the whole data analysis process. The detailed codebook, which outlines the themes and codes, is included in ~\autoref{appendix:codebook}.

\section{Survey Results}  

\subsection{Instructors' Current Practices of, Familiarity with, and Attitudes of GenAI}
We first provide an overview of instructors' familiarity with GenAI-related terms, their current use of GenAI in teaching and plans for future use, their current engagement with various instructional tasks with GenAI, and their attitudes towards receiving training with GenAI. Most of our survey respondents reported that they were very familiar (29.8\%) or somewhat familiar (52.8\%) with the concept of GenAI.

\begin{figure*}[h!]
\centering
\includegraphics[width=.7\linewidth]{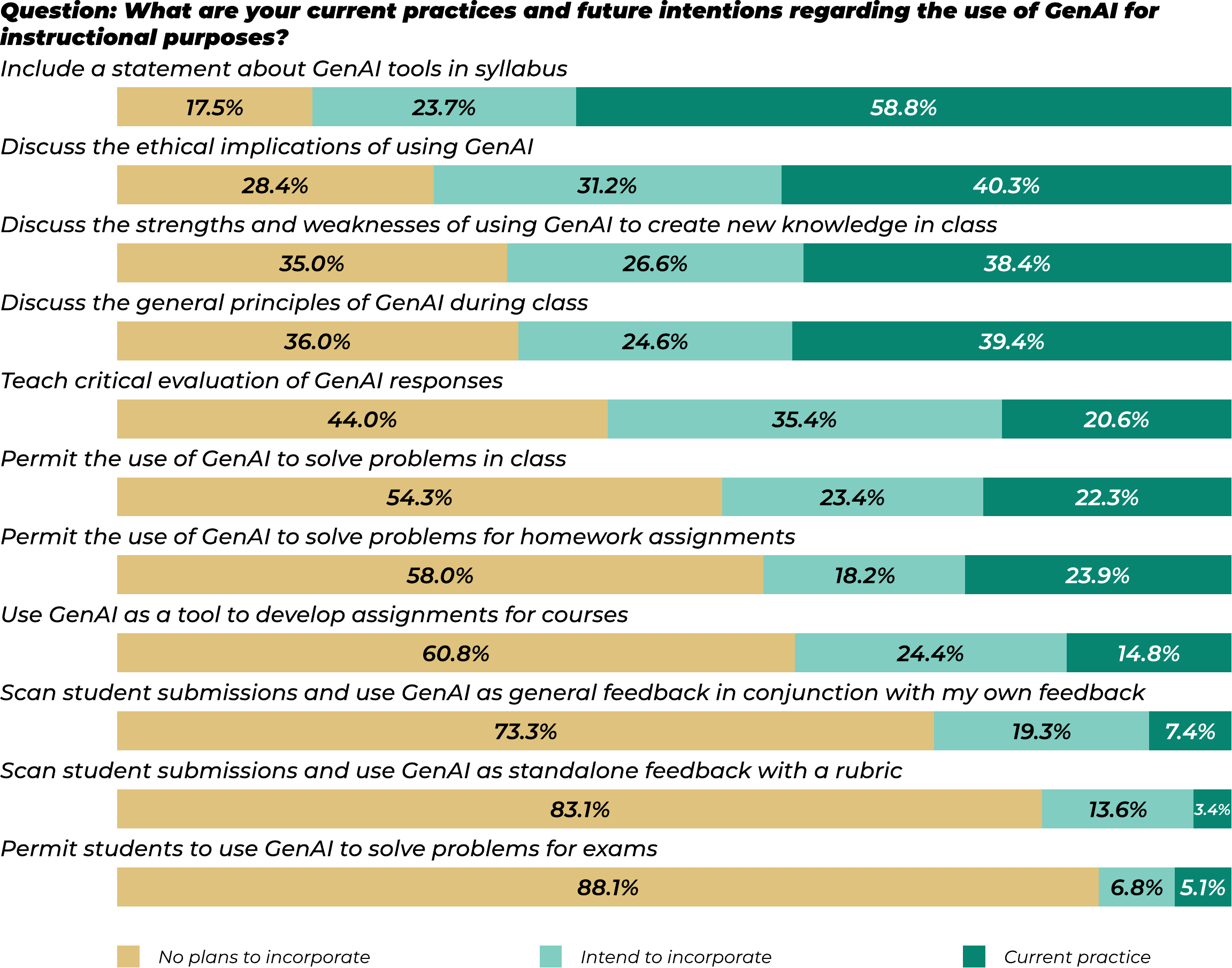}
\caption{Response distribution of current practices and future intentions regarding the use of GenAI tools for instruction, ordered by the percentage of respondents indicating `No plans to incorporate,' in ascending order.}
\label{fig:current_practices_and_future_intentions}
\end{figure*}

\textbf{Instructors' current practices and future intentions for GenAI use: }
\autoref{fig:current_practices_and_future_intentions} illustrates instructors' current practices and future intentions to incorporate GenAI into higher education. In our study, more than half of the instructors (58.8\%) had included a statement about GenAI tools in their syllabus. Most instructors were currently practicing or had a high intention of incorporating GenAI in activities, such as discussing the ethical implications of GenAI (40.3\%), the strengths and weaknesses of using GenAI to create new knowledge (38.4\%), the general principles of GenAI (39.4\%) in class. However, a significant majority (88.1\%) of instructors in our study did not currently permit students to use GenAI during exams. 

\begin{figure*}[h!]
\centering
\includegraphics[width=.8\linewidth]{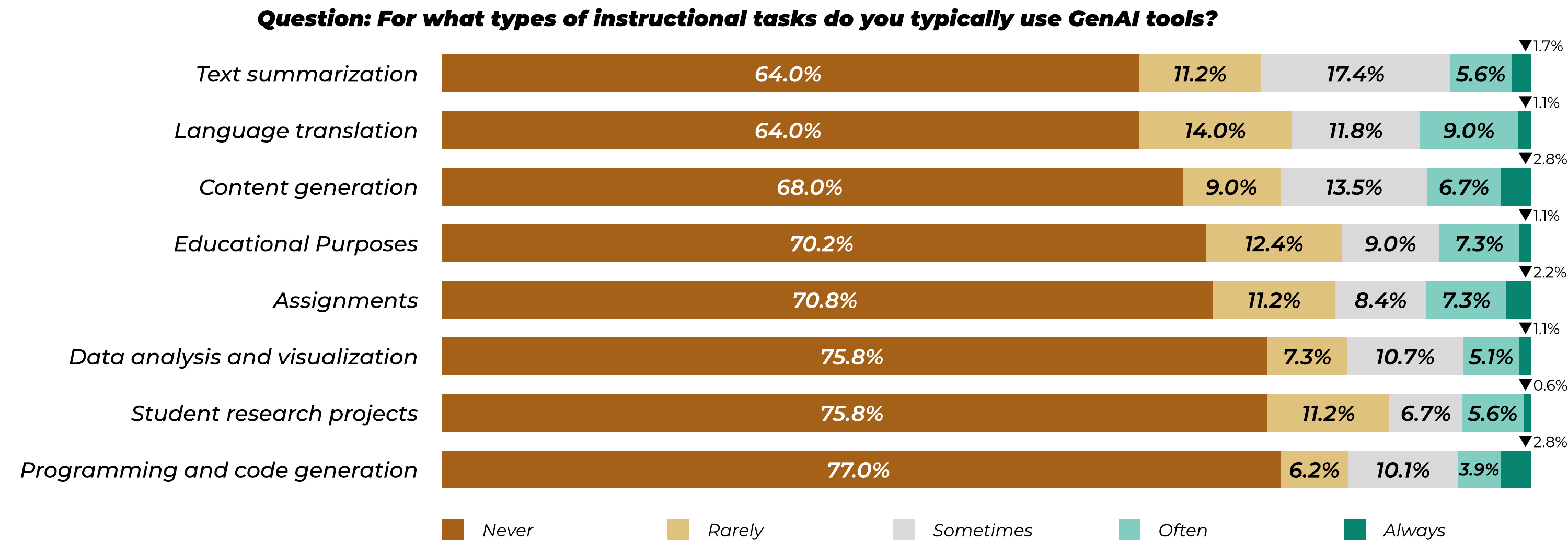}
\caption{Response distribution of the types of instructional tasks for which instructors typically use GenAI tools, ordered by the percentage of respondents indicating `Never,' in ascending order. For statements with the same percentage of `Never,' ties are broken by the percentage of `Rarely,' also in ascending order.}
\label{fig:instructional tasks}
\end{figure*}

\textbf{Instructors' engagement in instructional tasks with GenAI: }
Our results show the extent of their engagement in instructional tasks with various functionalities of GenAI in \autoref{fig:instructional tasks}. In our study, we can see that more than half of (from 64.0\% to 75.8\%) the instructors had never used GenAI tools in educational tasks, with some reporting frequent (from 6.7\% to 17.4\%) or constant (from 3.9\% to 9.0\%) use. Specifically, only up to 2.8\% instructors use GenAI as a routine for tasks of text and code generation. 

\begin{figure*}[h!]
\centering
\includegraphics[width=.78\linewidth]{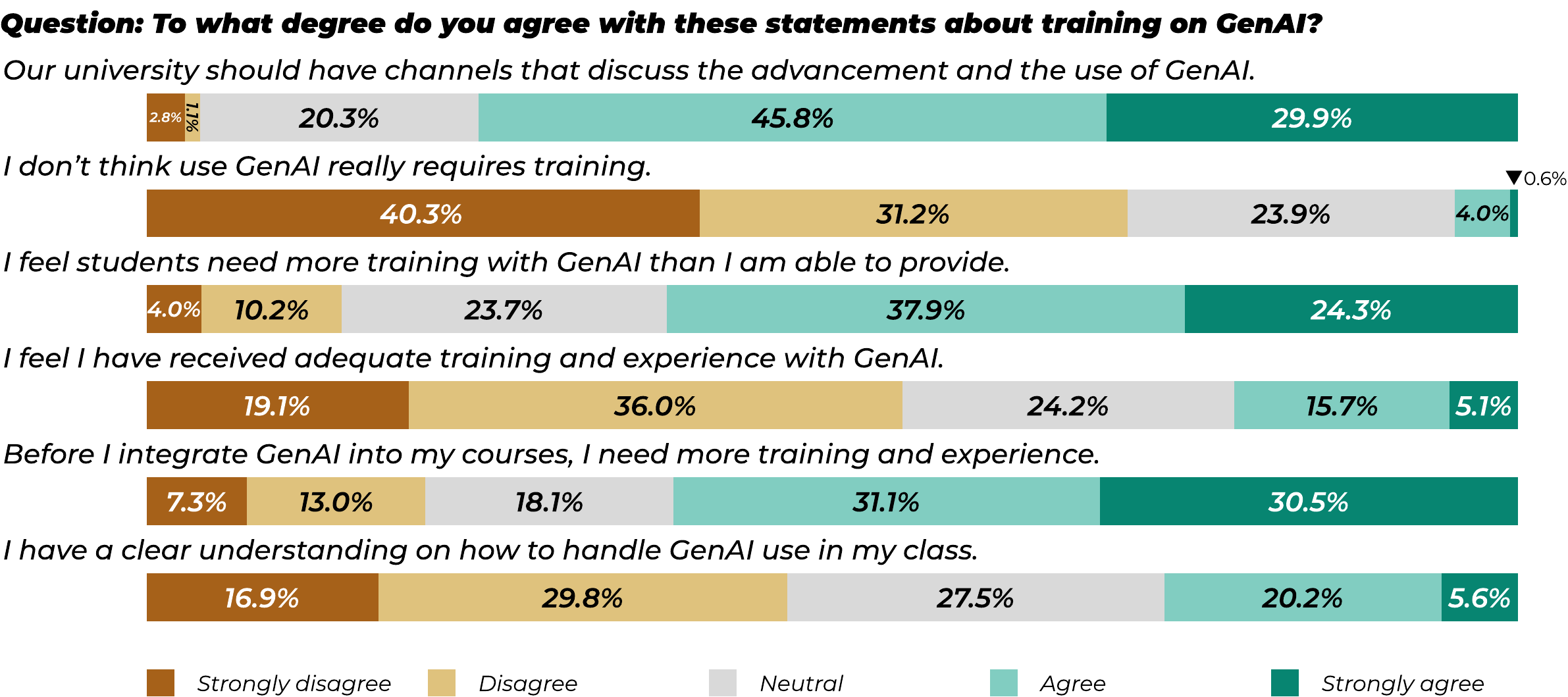}
\caption{Instructors’ attitudes towards training in GenAI, ordered by the percentage of responses in the most frequent category for each statement.}
\label{fig:new-Training}
\end{figure*}

\textbf{Instructors' attitudes towards receiving training with GenAI: }
We also asked about the instructors' attitudes towards receiving training with GenAI. As shown in \autoref{fig:new-Training}, the instructors in our study believed that the university (75.7$\%$), students (62.2$\%$), and themselves (61.6$\%$) needed more additional training and experience to effectively use GenAI. 46.7$\%$ of instructors felt they did not yet have a clear understanding of how to handle GenAI use in their class, and 55.1$\%$ considered their training and experience with GenAI to be insufficient.

\subsection{Dynamics of Trust and Distrust in GenAI among Instructors}
In addition to the overall understanding of instructors' practices of and experiences with GenAI, we further examine the trust and distrust instructors held toward GenAI in educational contexts. In this subsection, we explore the relationship between trust and distrust and how their perceptions shape their practices.

\subsubsection{Validity and Reliability of Trust and Distrust Scales} 

\begin{table}[h!]
\centering
\small
\caption{Factor Analysis results on trust and distrust questions, indicating effective identification of different underlying dimensions of trust and distrust in GenAI among instructors. Corresponding statements can be found in \autoref{tab:trust_distrust_questions_list}.}

\begin{tabular}{lcc}
    \toprule
    \textbf{Variable} & \textbf{Factor 1 (Trust)} & \textbf{Factor 2 (Distrust)} \\
    \midrule
    \textit{Trust}: Competence  & 0.813 &  \\
    \textit{Trust}: Personalization & 0.838 &  \\
    \textit{Trust}: Adaptability & 0.886 &  \\
    \textit{Trust}: Anticipation & 0.850 & \\
    \textit{Trust}: Transformation & 0.857 & \\
    \textit{Trust}: Enrichment & 0.722 & \\
    \midrule
    \textit{Distrust}: Malevolence &  & 0.640 \\
    \textit{Distrust}: Dishonesty &  & 0.761 \\
    \textit{Distrust}: Skepticism &  & 0.744 \\
    \textit{Distrust}: Inaccuracy &  & 0.830 \\
    \textit{Distrust}: Restriction &  & 0.690 \\
    \textit{Distrust}: Demotivation &  & 0.745 \\
    \bottomrule
\end{tabular}
\label{tab:factor_analysis}
\end{table}

To assess the validity and reliability of our scales, we conducted a factor analysis and calculated Cronbach’s $\alpha$ in SPSS. The value of the Kaiser–Meyer–Olkin test is 0.876, suggesting that our data is suitable for factor analysis. The result of Bartlett’s test of Sphericity was significant ($\chi^2$ =1364.879, df = 66, p< .01), indicating a significant correlation between the constructs in our data. \autoref{tab:factor_analysis} presents factor loadings for two factors. The factor loadings of six trust measurements range from 0.722 to 0.886, and the factor loadings of six distrust measurements range from 0.640 to 0.830. The total variation explained is 67.843$\%$, which is acceptable. Moreover, we achieved a good Cronbach’s $\alpha$ for trust (0.924) and Cronbach’s $\alpha$ for distrust (0.860).

\begin{figure}[h!]
\centering
\includegraphics[width=1\linewidth]{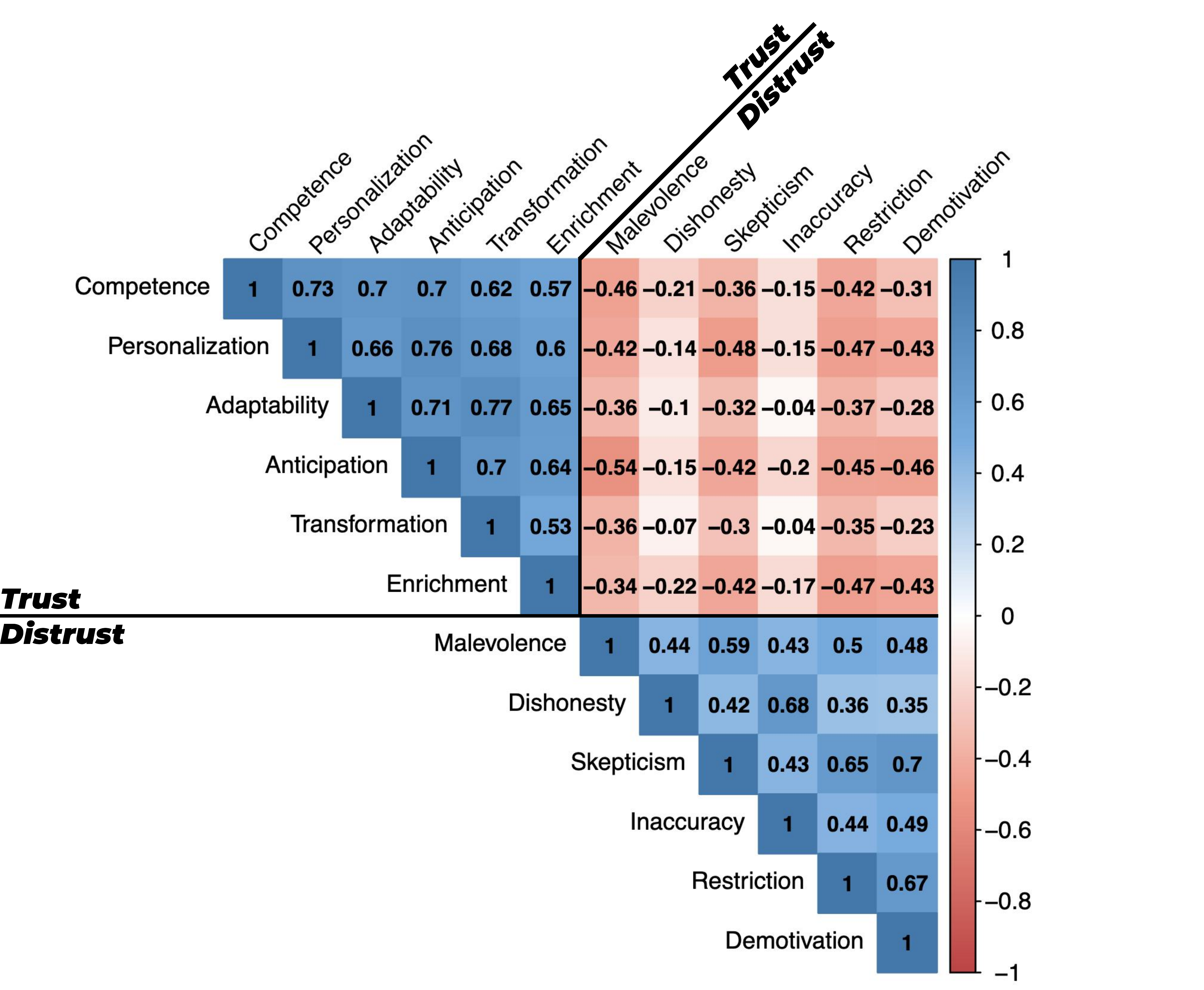}
\caption{Correlation matrix between all items for the trust and distrust scales to show how various trust (e.g., competence, personalization, etc.) and distrust (e.g., malevolence, dishonesty, etc.) attributes were statistically related. Each cell shows the correlation coefficient between two attributes, with values closer to 1 indicating a strong positive correlation and values closer to -1 indicating a strong negative correlation. The color gradient from blue to white represents increasing positive correlations, with blue denoting higher positive values and white indicating no or negative correlation. }
\label{fig:correlation graph}
\end{figure}

Our pairwise correlation graph (see \autoref{fig:correlation graph}) shows the correlations between trust and distrust items. The different items of trust are positively correlated with each other, as are the distinct items of distrust. However, trust and distrust are negatively correlated with each other. \textbf{Our results collectively suggest a delineation between trust and distrust, indicating that trust and distrust are related but distinct concepts that may co-exist in our study.} 

\subsubsection{Trust and Distrust May Co-exist} 
\label{subsubsec:trust_distrust_distribution}

\begin{figure}[h!]
\centering
\includegraphics[width=1\linewidth]{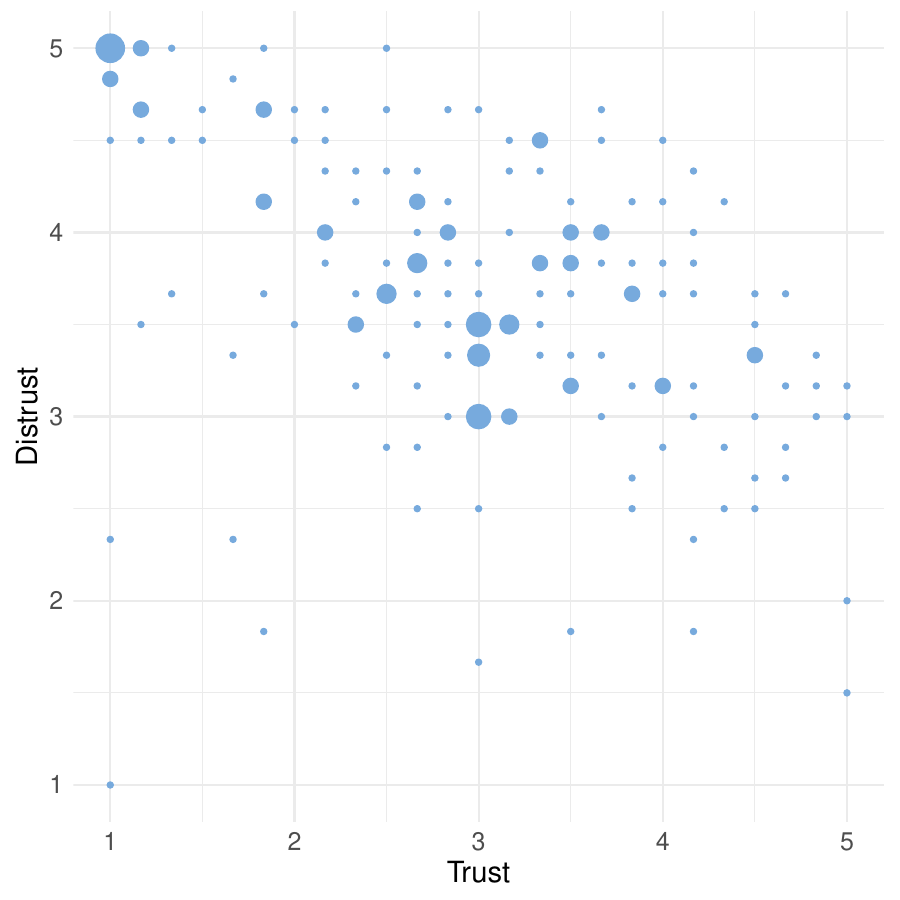}
\caption{Trust and distrust distribution. Each point on the plot represents an instructor's score for trust (X-axis) and distrust (Y-axis), on a scale from 1 to 5.}
\label{fig:Trust and distrust distribution}
\end{figure}

To further examine the relationship and interplay between trust and distrust among instructors towards GenAI, we plotted \autoref{fig:Trust and distrust distribution}. We conceptually separate instructors into four groups based on the median values of trust and distrust: High-Trust-High-Distrust (H-T-H-D), High-Trust-Low-Distrust (H-T-L-D), Low-Trust-High-Distrust (L-T-H-D), and Low-Trust-Low-Distrust (L-T-L-D). 
The low-trust-high-distrust group implies that the instructors had a correspondingly low trust in the use of GenAI while having a high distrust, whereas those in the high-trust-high-distrust group imply that the instructors have high distrust in the use of GenAI along with high trust. An absence of respondents with extreme levels of distrust was observed (i.e., the lower sections of the distrust axis are much less populated). 

These results suggest a possible coexistence of trust and distrust among instructors. Our observation prompts two questions: (i) \textit{How do the dynamics of trust and distrust manifest within different educational contexts?} (ii) \textit{How do trust and distrust become manifested among instructors in the educational context?} For the first question, we provided an analysis in \autoref{subsubsection:4.2.3 section} to help explore the variations of the dynamics of trust and distrust. The second question motivated us to further investigate through the following qualitative analysis in \autoref{sec:qual_findings}.

\subsubsection{Variations in Trust and Distrust Towards GenAI Among Instructors}
\label{subsubsection:4.2.3 section} 
We then explore potential variations in trust and distrust toward GenAI in educational contexts among our participants, such as familiarity with GenAI. We hypothesize that instructors' trust and distrust levels might be associated with their familiarity with GenAI. The level of familiarity with the GenAI concept may indicate that instructors are informed about GenAI, with some understanding of GenAI's potential and limitations. Essentially, understanding GenAI's capabilities and limitations allows instructors to make informed decisions about its use. For example, if instructors possess limited knowledge about GenAI, they may either exhibit excessive trust without acknowledging associated risks---termed ``blind trust''~\cite{klingbeil2024trust}—or they may display distrust due to their ignorance of the GenAI's potential benefits, resulting in ``blind distrust.''~\cite{benamati2010productive}. Both blind trust and blind distrust can be harmful to the calibrated trust building~\cite{benamati2010productive}.

\begin{figure}[h!]
\centering
\includegraphics[width=1\linewidth]{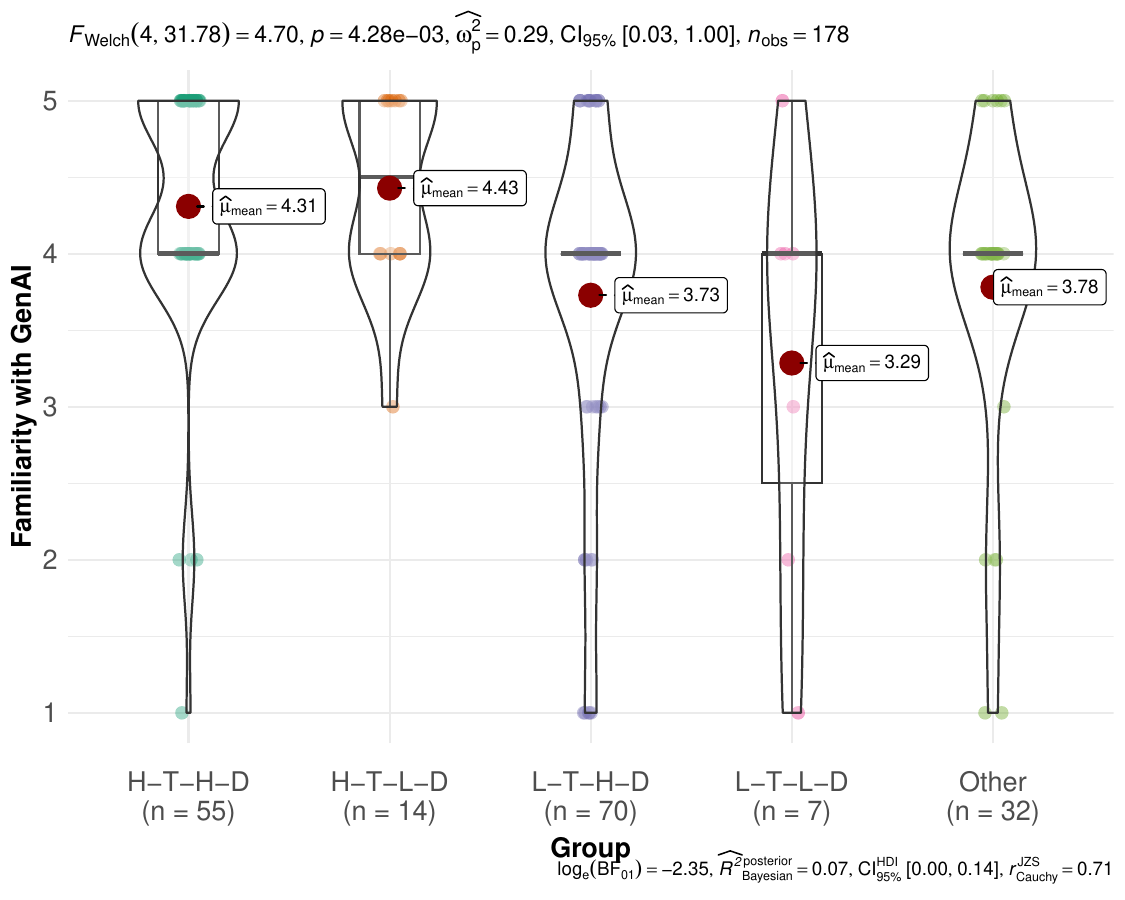}
\caption{Differences in familiarity with GenAI across different groups.}
\label{fig:Group_familiarity}
\end{figure}

\autoref{fig:Group_familiarity} shows that there are statistically significant differences in GenAI familiarity across the trust-distrust groups (\emph{p} < 0.05), with a moderate to large effect of group categorization on the levels of familiarity with GenAI. To break down the groups and familiarity levels, we can see that: \\
(1) \textbf{High Trust with Higher Familiarity:} Groups with higher trust levels 
(\highTrustHighDistrust, \highTrustLowDistrust)
tend to exhibit higher familiarity. This might indicate that building high trust among instructors may help them become more familiar with GenAI's capabilities and limitations, even if it is accompanied by high distrust in some cases. \\
(2) \textbf{Mixed Trust and Distrust with Low-to-Moderate Familiarity: } Varied levels of trust and distrust groups 
(\lowTrustHighDistrust, \lowTrustLowDistrust)
show different familiarity, with the former presenting moderate familiarity and the latter showing lower familiarity. This might suggest that instructors with limited knowledge of GenAI could potentially exhibit mixed trust and distrust and even cautious or negative perceptions of GenAI's role in educational settings.

\subsubsection{Trust and Distrust in GenAI Across Teaching Levels}
We also explored the relationship between trust/distrust and teaching levels (undergraduate and graduate levels), given our consideration of teaching requirements and pedagogical approaches may vary between undergraduate and graduate education, which might impact instructors' perceptions and usage of GenAI.

\begin{figure*}[h!]
\centering
\includegraphics[width=1\linewidth]{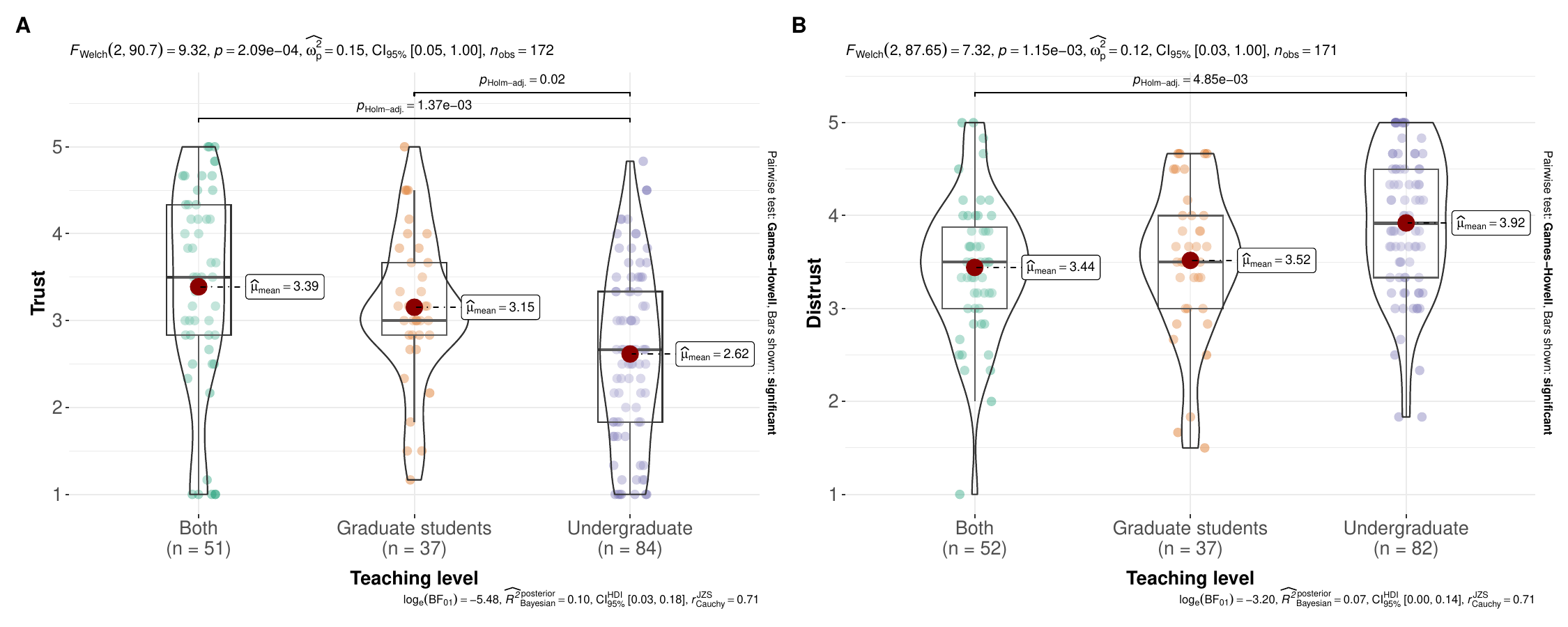}
\caption{Differences in \textbf{(A)} trust and \textbf{(B)} distrust across different teaching levels (undergraduate vs. graduate level).}
\label{fig:Different teaching level}
\end{figure*}

We conducted Welch's ANOVA analysis to examine instructors' differences in trust and distrust across different teaching levels. As shown in \autoref{fig:Different teaching level}, we found the educational levels at which instructors teach (undergraduate vs. graduate) significantly influence their trust and distrust in GenAI technologies (\emph{p} < 0.05). 

In terms of trust, as shown in \autoref{fig:Different teaching level} (A), our results show that instructors teaching both undergraduate and graduate levels show the highest mean trust level at 3.39, suggesting they are generally more trusting GenAI. Instructors teaching only undergraduate students show the lowest mean trust at 2.62, suggesting they are more cautious or skeptical about the benefits or applicability of GenAI in their teaching. Those teaching exclusively graduate students show a moderately high mean trust at 3.15, indicating an acceptable but slightly less enthusiastic compared to instructors teaching both levels.

In terms of distrust, as shown in \autoref{fig:Different teaching level} (B), we found that graduate-level instructors show slightly lower distrust; instructors at the undergraduate level show the highest distrust, whereas instructors teaching both levels are shown to have an intermediate level of trust. This is roughly the opposite trend of the data distribution of trust values, although the difference between the means of instructors teaching both levels and graduate-level groups is quite small relative to trust. In addition, the data distribution of distrust among instructors in \autoref{fig:Different teaching level} (B) is denser compared to that of trust in \autoref{fig:Different teaching level} (A), indicating that instructors' distrust tended to be more widespread or intense.


\takeaway{Our findings reveal a gap between instructors' familiarity with GenAI concepts and their actual use of GenAI for direct instructional tasks. While many instructors in our study include discussions about GenAI's ethical implications and general principles in their syllabus and classroom activities and report moderate to high familiarity with GenAI concepts, few have incorporated GenAI into hands-on instructional tasks; their practical experience of GenAI in teaching remains limited. Additionally, although instructors recognize the potential benefits of GenAI in enhancing education, many still expressed the need for more training and clearer guidelines on how to integrate these tools into higher education effectively. 

Our results also show that trust and distrust are related but distinct concepts that may co-exist for instructors in the context of higher education. We find that greater familiarity with GenAI exhibited higher trust among four trust and distrust groups of instructors (\highTrustHighDistrust, \highTrustLowDistrust, \lowTrustHighDistrust, \lowTrustLowDistrust), indicating that a higher trust tends to enhance the understanding of GenAI. In contrast, increased distrust was associated with lower familiarity with GenAI, highlighting the need for more systematic guidance and training to boost confidence in GenAI's capabilities for instructors. We also found significant differences in instructors' trust and distrust across various teaching levels. Specifically, instructors who solely taught undergraduate courses showed lower trust compared to those who taught both undergraduate and graduate courses; distrust was highest among instructors teaching only undergraduate courses.}


\section{Qualitative Findings}
\label{sec:qual_findings}
Our qualitative analysis of open-ended survey questions also provides insights into how trust and distrust in GenAI manifest among instructors. Below, we present different themes of instructors' trust and distrust in GenAI for teaching and learning (also see Table~\ref{tab:qual_findings_summary} for an overview of themes).

\subsection{Distrust in GenAI for Teaching and Learning}
Distrust in GenAI often arises from concerns about the limitations, potential inaccuracies, and the broader implications of GenAI's deployment in educational settings~\cite{michel2023challenges}. Below, we describe the ways in which distrust in GenAI became manifested among instructors in our study.

\subsubsection{Blind Distrust in GenAI} \label{subsubsection:blind_distrust}  Some instructors in our study \textbf{expressed significant distrust toward GenAI, despite lacking direct experience with it or providing specific reasons for their reservations, regardless of their familiarity with GenAI.} Overall, when triangulating our survey and interview data, we found that instructors in our low-trust-high-distrust group have moderate level of familiarity with GenAI (average familiarity of GenAI concept of 3.86, compared to the overall average of 3.96 across all groups), but have shown extensive distrust towards GenAI without clear reasons indicated. For example, when asked about any positive examples of their GenAI practice, P31 (\lowTrustHighDistrust) stated:
\begin{quote}
    \emph{``There are none [positive examples of GenAI]. Generative AI does not belong here and we absolutely should not be integrating it.'' (P31, \lowTrustHighDistrust)}
\end{quote}
P31's quote indicates a sense of hostility towards GenAI, as indicated by phrases like ``\textit{absolutely should not}'' suggests a strong and non-negotiable opposition, reflecting deeper concerns or fears about the impact of GenAI on education. And yet, P31's quote also shows that no reasons were given to justify this sentiment, indicating a sense of ``blind distrust''.      

Similarly, P79 (\lowTrustHighDistrust) said \emph{``I forbid its use in my classes''} even though \emph{``I have no experience with it''}. Likewise, P16 (\lowTrustHighDistrust) conveyed their complete rejection of GenAI, stating: \emph{``I am opposed entirely to the use of Gen AI, full stop.'' }
These responses illustrate a phenomenon of blind distrust among instructors, where they reject GenAI outright without any hands-on experience or clear justification.

\subsubsection{Concerns surrounding Social Justice Issues}
\label{subsubsec:social_justice}
Instructors raised \textbf{concerns about the content generated by GenAI often contains inherent biases related to linguistic justice, social justice, and copyright violations.} Many instructors in our study were worried about how GenAI might reinforce existing biases and perpetuate societal inequalities. These biases result not only from the content output but also from the broader implications of how GenAI is deployed and utilized in educational contexts. 
P132 (\lowTrustHighDistrust) highlighted the potential loss of linguistic diversity and the reinforcement of mainstream norms regarding gender, race, and class that may arise from GenAI, potentially marginalizing groups that do not conform to these standards.
\begin{quote}
    \emph{``We are very concerned about the issue of linguistic justice and how these platforms erase differences of dialect and other variants of English. Using generative AI in writing for revision to make writing `more assertive' or `more professional' can also reinforce dominant norms about gender, race, and class...'' (P132, \lowTrustHighDistrust)}
\end{quote}

Meanwhile, the very technologies intended to enhance learning could instead widen existing educational gaps. As P57(\highTrustHighDistrust) said, 
\begin{quote}
    \emph{``The students who are struggling and have limited time are the ones more likely to use it inappropriately and also more likely to suffer negative consequences. Hence, using current iterations introduces a social justice issue--the people who need the most help are the ones who will suffer the most costs of its use.'' (P57, \highTrustHighDistrust)} 
\end{quote}
P57 highlighted the latent risks associated with the practical application of GenAI, particularly for students who may be poorly equipped, time-constrained, and prone to misuse the technology, exposing them to more adverse impacts. P57's observations raised critical social justice concerns, indicating that GenAI might disproportionately benefit those already well-equipped and resource-rich, further marginalizing those in need of the most support. 

P8 (\lowTrustHighDistrust) pointed out that the exceptional performance of GenAI may come at the expense of \emph{``exploiting workers in the training process''}. Raising concerns of data annotation specialists, whose labor protections and rights might be inadequate, facing not only the physical strain but also the mental anguish from exposure to toxic and misleading information.

Moreover, GenAI generates content, such as text and images, by training large models on comprehensive datasets consisting of real data and existing creative works. This process, which includes but is not limited to language and image models, raises potential copyright infringement issues due to the breadth of copyrighted material used in the training datasets. As reported by P140 (\highTrustHighDistrust), \emph{``Everything we see, hear, view, read has POV and bias. One must be vigilant. One must not charge money or claim to have one's own products generated by AI. Copyrights, fair use, etc....''} P19 (\lowTrustHighDistrust) also described this issue: \emph{``Among other things, the models were built by infringing the intellectual property of scholars and artists including faculty on our own campus.''}

\subsubsection{Concerns about Environmental and Sustainability}
\label{subsubsec:sustainability}
\textbf{GenAI, owing to its demand for substantial resources, may exert a potential negative impact on the environment.} P8 (\lowTrustHighDistrust) noted that \emph{``Teaching everyone a better understanding of what these models actually can and cannot do, as well as of their hidden costs (... high energy and water costs), would be very valuable.''} Similarly, P45 (\lowTrustHighDistrust) highlighted the significant resources required to operate such technologies and questioned their sustainability:
\begin{quote}
    \emph{``The social and environmental impacts of using GenAI are already well documented. Given the energy and water uses of GenAI, how can we justify using them in the classroom? Until these are fixed, using GenAI is detrimental to humans and the environment. There is no way around this.'' (P45, \lowTrustHighDistrust)}
\end{quote}

P20 (low-trust-high-distrust) believed that the environmental detriments associated with GenAI were also manifested on campus, particularly as there was an increased reliance on the use of GenAI among the general public in educational settings. 

\begin{quote}
    \emph{``I also think, as a campus, we need to grapple with the massive environmental toll of Generative AI. We tout our excellent green initiatives and are also throwing ourselves into AI, which relies on a truly massive amount of energy and water consumption. '' (P20, low-trust-high-distrust)}
\end{quote}

Collectively, these quotes indicate a significant level of concern about the broader environmental and ethical implications of GenAI use. Instructors highlighted the need for careful consideration of both the direct and indirect effects of integrating such technologies into educational practices, urging a more ethically and environmentally conscious approach. 

\subsubsection{Ethical and Pedagogical Concerns} 
In addition to the concerns regarding social and environmental implications, instructors in our study also raised issues regarding the pedagogical and ethical dimensions of using GenAI within educational settings. 

\textbf{Instructors expressed concerns about students' misuse of GenAI in terms of academic integrity and the learning process.} For example, P22 (\highTrustHighDistrust) described an instance where a student appeared to be overly reliant on AI to complete a project, incorporating fictitious sources and data analysis techniques.
\begin{quote}
    \emph{``I came across a student who seemed like she heavily relied on AI for a project. She had some fictional sources in her project along. Additionally, she mentioned data analysis techniques that were not mentioned in our course.'' (P22, \highTrustHighDistrust)}
\end{quote}

Likewise, P57 (\highTrustHighDistrust) also expressed concerns about the deceptive behavior of students attributing GenAI content to their own results: \emph{``A minority of students will use this regularly and claim the work is their own ... and get in the habit of cheating throughout College and then into their professional lives.”} P117 (\lowTrustHighDistrust) unequivocally labeled the use of GenAI for academic tasks as cheating, advocating for genuine learning: \emph{``It is cheating. I also think that we, as an institution, should be above this and should stand up for actual learning (and pilfering copyrighted words, no less).''} 

These plagiarism concerns have led educators to contemplate reforming their course syllabi and assessment strategies. Specifically, P40 (\lowTrustHighDistrust) said, \emph{``If I can't trust that work done outside class was actually done by the students and not an AI, I'll have to rely more on in-class exams to assess student learning.''} Likewise, P124 (\highTrustHighDistrust) questioned the meaning of assignments in the era of GenAI, \emph{``If AI can do the assignment, what is the point of giving that assignment?''}

Beyond the integrity issues, instructors in our study \textbf{raised a lot of pedagogical concerns about GenAI potentially undermining the development of critical thinking and analytical skills.} They were concerned that students might become overly reliant on the content provided by AI without the necessary questioning, which could lead to limitations in students' cognitive abilities and creative thinking. For instance, P54 (\highTrustHighDistrust) explained the significance of critical thinking and learning:
\begin{quote}
    \emph{``It seems that the tools function best in education with highly motivated students who have well-developed critical thinking skills. I'm concerned that the typical student will only be harmed by the perceived shortcuts of generative AI.'' (P54, \highTrustHighDistrust)}
\end{quote}

P105 (\highTrustLowDistrust) expressed concerns about a potential decline in students’ comprehension skills due to the overly complex content provided by GenAI: \emph{``My students do not understand the content provided by generative AI. ChatGPT is giving too advanced solutions.''} This emphasized that students should actually learn and understand knowledge rather than simply using GenAI as a convenient tool for completing tasks. In addition to comprehension skills, P77 (\lowTrustHighDistrust) also noted the impact of GenAI on students' writing skills: \emph{``I believe that language is the tangible expression of thought. Students who rely on Gen AI for writing assignments may compromise their ability to write well and think clearly.''} P177 (\highTrustHighDistrust) raised a critical pedagogical concern: the potential for GenAI to undermine essential educational goals such as seeking and understanding information during learning. By potentially offering ready-made answers, GenAI could detract from the process of critical inquiry and evidence-based reasoning that is fundamental in disciplines like public policy. P177 mentioned:
\begin{quote}
\emph{``So far, in my Public Policy course, I have had no reason to use GenAI, and it would actually be counterproductive to my efforts to teach students to better analyze where they get their information and why they believe what they believe (as future public policy practitioners).'' (P177, \highTrustHighDistrust)}
\end{quote}

Instructors further considered the \textbf{broader homogenization effects of deploying GenAI.} 
In particular, P24 (\lowTrustHighDistrust) stated that: \emph{``Homogenization of society! in terms of views, writing style, hierarchy of problems, etc.''} Similarly, P46 (\highTrustLowDistrust) expressed their concern of \emph{``Losing the individual's voice.''} These quotes indicate that as teachers and students widely use GenAI, their thought processes may gradually be influenced by it, weakening critical thinking and, thus, the emergence of homogeneous behaviors amid the fact that GenAI is not creating any \textit{new} content~\cite{feuerriegel2024generative}. 

P137 (\lowTrustHighDistrust) also stated the homogenization of knowledge, believing that students' works were becoming progressively more homogenized in content and had clear traces of GenAI, as explained: \emph{``Students don't learn how to read or how to find evidence. They turn in work that says the same banal things over and over again.''} P9 (\highTrustHighDistrust) noted the same issue: \emph{``I have had short writing assignments in my courses, and when I grade them I see the same language over and over again (ChatGPT likes to delve and use the verb boast).''} 



These findings reflect deep pedagogical concerns about the broader educational implications of GenAI, underscoring the need to carefully consider how these tools are integrated into learning environments to support rather than hinder student development.

\subsubsection{Additional Burdens on Faculty}
It is also important to recognize how \textbf{GenAI poses challenges that extend faculty responsibilities} as we consider its broader impact on students' learning, particularly concerning academic integrity and critical thinking. Instructors in our study explained that their concern about student reliance on GenAI leads to the additional burdens faced by instructors, who must adapt their assessment methods and manage increased workloads to maintain academic standards. P20 (\lowTrustHighDistrust) explained,
\begin{quote}
    \emph{``I hate that I now have to spend additional time grading, trying to test whether written assignments were generated with AI. The built-in plagiarism software in Blackboard is already hot garbage ... This adds time and burden to what everyone already agrees is the worst part of teaching.'' (P20, \lowTrustHighDistrust)}
\end{quote}

The challenge of fair assessment is exacerbated by the varied ability of students to use GenAI tools effectively, as noted by P149 (\lowTrustHighDistrust):
\begin{quote}
    \emph{``Inability to evaluate students fairly when GenAI tools are used; as students do not receive any institutional training/education on how to effectively use these tools, it means that some students will be naturally adept at using them while others will fall behind due to lack of access/training/practice.'' (P149, \lowTrustHighDistrust)}
\end{quote}

These insights suggest the additional responsibilities and challenges that instructors face as GenAI becomes more prevalent in academic settings, highlighting the need for institutional support and policy adjustments to address these emerging issues effectively. However, the concerns remain among educators that institutional policy could potentially influence their current practice, as P53 (\highTrustHighDistrust) mentioned, \emph{``I actually worry that the efficiencies it creates will allow administrators to put larger numbers of students in classes, create expectations about AI feedback on assignments, and reduce the real connection between faculty and students.''} P63 (\highTrustHighDistrust) articulated that any constraints placed on GenAI by institutions could inadvertently result in covert discrimination against individuals with disabilities:   

\begin{quote}
    \emph{``I worry that faculty, staff and students with disabilities are being told on certain university syllabus that they can't use AI when that is how they got to the university in the first place! It would be distressing if the need to use AI were somehow made more difficult to obtain, like Zoom was before the pandemic.'' (P63, \highTrustHighDistrust)} 
\end{quote}

\subsection{Trust in GenAI for Teaching and Learning}
In addition to various ways of distrusting GenAI among our participants, we also discuss how instructors in our study elaborate on their trust formation. 

\subsubsection{Believing GenAI in Transforming Educational Practices}
\textbf{High trust in GenAI among some instructors in our study is characterized by a strong belief in its essential role and transformative potential within educational practices.} Participants who held high trust emphasized GenAI's substantial advantages and viewed its integration as critical to the evolution of education. For example, participant P174 (\highTrustLowDistrust) compared the emergence of GenAI to other important technological advancements:
\begin{quote}
    \emph{``My major interaction with students is in the context of advising them on their graduate school research. Generative AI is a revolutionary change akin to the wide availability of libraries, computers, and the internet for exploring new topics. Not adapting to use AI to its full potential is as misguided as not using libraries, computers, or the internet.'' (P174, \highTrustLowDistrust)}
\end{quote}
The quote from P174 indicates great confidence in the transformative capabilities of GenAI, equating its impact on education to that of historical technological advancements, such as libraries, computers, and the internet. This comparison suggests a strong belief in the necessity and inevitability of integrating GenAI into educational practices.

\subsubsection{``Trust, but Verify''}
While high trust among our participants indicates their enthusiasm for the broad adoption of GenAI, many \textbf{ adopted a \textit{``trust, but verify''} approach in using GenAI for instructional tasks.} For example, P57 (\highTrustHighDistrust) described a balanced use of GenAI for data gathering and verification to increase accuracy:
\begin{quote}
    \emph{``I have asked ChatGPT to collate simple factual data for some topics, such as summarizing demographic information from census data, and used that in class. I have verified the accuracy of some requests (sub-sampled) to check the accuracy and for many simple questions, ChatGPT does a good job.'' } (P57, \highTrustHighDistrust) 
\end{quote}
P57's quote suggests a sense of calibrated trust involving a thoughtful and balanced approach to integrating GenAI in educational settings, where instructors carefully consider both the benefits and limitations of GenAI.

In addition to content verification, seeking additional resources to supplement learning can be seen as another manifestation of calibrated trust. P140 (\highTrustHighDistrust) commented:
\begin{quote}
    \emph{``When we plugged in a Lesson Plan prompt, AI gave us a `skeleton' plan that kickstarted our thinking and planning--but we still had to gather the resources for students to use to learn.'' (P140, \highTrustHighDistrust)} 
\end{quote} 
P140 described a practice in which instructors did not fully use the content provided by GenAI; instead, they used it as a reference and made improvements based on it. This adaptive strategy was especially beneficial in educational contexts, where the extensive teaching responsibilities can be mitigated by effectively harnessing GenAI enhancements.

\subsubsection{Guiding Students to Critically Engage with GenAI to Build Calibrated Trust}

\textbf{Instructors also play an important role in guiding students on how to effectively use GenAI to embrace a judicious and balanced approach to enhancing learning outcomes.} They encourage students to critically engage with GenAI, emphasizing its role as a supplementary tool for learning enhancement rather than a replacement for traditional educational methods. For example, P108 (\highTrustLowDistrust) highlighted how their students effectively adapted GenAI to accomplish tasks, yet they still required manual intervention to improve the accuracy and quality of the results:
\begin{quote}
    \emph{``My students are currently using GenAI as a part of their work to summarize thousands of PDF documents (reports).  In order to assess how well the GenAI model is working, for a subset of the reports, they have also done the process manually.''  (P108, \highTrustLowDistrust)} 
\end{quote} 
While students independently used GenAI appropriately on their own, instructors in our study also actively took steps to cultivate a calibrated trust in GenAI to support their learning. For example, P161 (\highTrustHighDistrust) described their practice of encouraging students to customize GenAI-powered applications to facilitate an ongoing optimization of their interactions: 
\begin{quote}
    \emph{``I asked my students to create a custom GPT to help them study for class. The results forced the students to think about what information the GPT needed to be helpful. It also forced the students to revise their queries until they got better results.''  P161 (\highTrustHighDistrust)}
\end{quote}
In P161's case, as students evaluated GenAI responses and refined their queries to achieve better outcomes, they also engaged in verifying these outcomes, thereby experiencing the process of calibrating their trust in the technology.

Moreover, P125 (\highTrustHighDistrust) pointed out that the precision of prompts is crucial when using GenAI, particularly in code generation, as vague or illogical inputs often result in irrelevant outputs. Students should thoroughly understand both the functions and the logic behind the generated code before effectively incorporating it into their projects: 
\begin{quote}
    \emph{``The GenAI tool did a great job of creating code, but the students needed to look up the functions to understand what was being done. Without understanding the suggested code, it was not possible for students to incorporate it into their work. Further, the writing of the prompt was actually the writing of the pseudocode, which requires students to organize their thoughts precisely. When pseudocode (prompt) is not specific (and detailed) it generates irrelevant code.'' (P125, \highTrustHighDistrust)}
\end{quote}

\takeaway{
Collectively, our qualitative findings reveal various manifestations of instructors' distrust of GenAI. Key concerns include the social justice implications of GenAI-generated content and the potential for a digital divide, where less-equipped, time-constrained students might misuse the technology. Instructors also expressed apprehensions about the environmental and sustainability issues linked to the significant resource demands of GenAI. Additionally, our participants expressed ethical and pedagogical concerns, as well as the additional burdens put on instructors, such as spending additional time on plagiarism detection for AI-generated content.

Turning to trust, our findings reveal that instructors in our study have shown high trust as they had a strong belief in the capabilities of GenAI in teaching and learning. Many instructors in our study adopted a ``trust, but verify'' approach, thoughtfully integrating GenAI to complement and enhance traditional teaching methodologies, which helped align with their pedagogical goals. Moreover, instructors play an essential role in helping their students build calibrated trust through practical engagement (e.g., creating customized GenAI-powered tools) to help students understand the capabilities and limitations of GenAI. 
}

\begin{table*}[ht]
\centering
\footnotesize
\caption{Summary of main themes regarding instructors' trust and distrust in GenAI}
\label{tab:qual_findings_summary}
\begin{tabular}{p{3.6cm} p{4.6cm} p{6cm}}
\toprule
\textbf{Theme} & \textbf{Description} & \textbf{Example Quote} \\
\midrule
\multicolumn{3}{l}{\textbf{Distrust in GenAI for Teaching and Learning}} \\
\cmidrule(lr){1-3}
\textbf{Blind distrust in GenAI} 
& Instructors expressing distrust without direct experience or explicit reasons 
& \emph{``Generative AI does not belong here and we absolutely should not be integrating it.''} (P31) \\
\addlinespace
\textbf{Concerns surrounding social justice issues} 
& Worries that GenAI may reinforce biases, inequities, and copyright issues 
& \emph{``We are very concerned about ... these platforms erase differences of dialect ... reinforce dominant norms about gender, race, and class.''} (P132) \\ 
\addlinespace
\textbf{Concerns about environmental and sustainability} 
& Awareness of the high resource consumption of GenAI 
& \emph{``Given the energy and water uses of GenAI, how can we justify using them in the classroom? ... using GenAI is detrimental to humans and environment.''} (P45) \\
\addlinespace
\textbf{Ethical and pedagogical concerns} 
& Fears of misuse (plagiarism, undermining critical thinking) and the homogenization of student work 
& \emph{``If AI can do the assignment, what is the point of giving that assignment?''} (P124) \\
\addlinespace
\textbf{Additional burdens on faculty}
& Instructors face extra workload in detection, grading, and accommodating policy changes 
& \emph{``I hate that I now have to spend additional time grading, trying to test whether written assignments were generated with AI.''} (P20) \\
\midrule
\multicolumn{3}{l}{\textbf{Trust in GenAI for Teaching and Learning}} \\
\cmidrule(lr){1-3}
\textbf{Believing GenAI in transforming educational practices} 
& High-trust instructors view GenAI as a transformative technology for education
& \emph{``Generative AI is a revolutionary change akin to the wide availability of libraries, computers, and the internet.''} (P174) \\
\addlinespace
\textbf{``Trust, but verify''} 
& Calibrated trust that involves verification of AI-generated content 
& \emph{``I have asked ChatGPT to collate simple factual data ... I have verified the accuracy ... for many simple questions, ChatGPT does a good job.''} (P57) \\ 
\addlinespace
\textbf{Guiding students to critically engage with GenAI} 
& Encouraging students to refine prompts, verify results, and integrate outputs thoughtfully 
& \emph{``I asked my students to create a custom GPT to help them study ... it forced the students to revise their queries until they got better results.''} (P161) \\
\bottomrule
\end{tabular}
\end{table*}

\section{Discussion} 
Our survey results revealed instructors' GenAI experiences and practices, as well as trust and distrust in GenAI for teaching and learning. Based on our findings, we discuss several design implications to support trust calibration in GenAI for teaching and learning among instructors and students, calling for more empirical studies and evidence-based adjustments to examine the future interventions on faculty trust in GenAI, as well as providing directions for institutional-level to enhance GenAI literacy among faculty.

\subsection{Instructor Practices of GenAI in Higher Education \&  Pedagogical Implications }
Our findings collectively suggest that while instructors have a reasonably high level of familiarity and positive intent toward GenAI in general (e.g., including it in their syllabi and discussing its ethical implications), their actual usage for everyday instructional tasks remains low. Tension emerges around exam settings, where a large majority of instructors still prohibit GenAI use, indicating concerns about academic integrity or uncertainty about how to govern AI-assisted assessment. Moreover, many instructors believe that both the institution and students need more structured support to develop GenAI competencies—yet nearly half feel they lack clarity on how to integrate GenAI effectively in class, revealing a gap between theoretical acceptance of GenAI’s potential and practical readiness to leverage it fully. Our results highlight an opportunity for targeted training, policy guidance, and best-practice sharing to help educators align their enthusiasm for GenAI’s transformative promise with concrete, effective classroom adoption.

Our findings also mirror those of other GenAI surveys in diverse cultural contexts. In Bulgaria, for example, approximately 40\% of surveyed instructors~\cite{kiryakova2023chatgpt} indicated they would use GenAI to create exercises and quizzes, similar to what we observed. Meanwhile, other U.S.-based studies~\cite{ghimire2024generative, amani2023generative} reported higher overall rates of GenAI use, potentially reflecting disciplinary differences: while our institution focuses primarily on the liberal arts, these other studies included more instructors from engineering and technical disciplines. As a result, the extent to which GenAI becomes a routine classroom tool may hinge on its alignment with a field’s pedagogical goals, suggesting that future research should explore how academic disciplines and institutional priorities affect instructors’ willingness and ability to incorporate GenAI.

Additionally, it is worth noting that most of the surveys to which we compared our findings were conducted in 2023, either explicitly stated or inferred through publication timelines. Given how rapidly GenAI continues to evolve, such timing differences may limit direct comparisons. We, therefore, recommend that future survey-based studies prominently report their data-collection dates (ideally in their abstracts or methods sections), to help researchers and readers contextualize results within the ever-shifting GenAI landscape. 

From a pedagogical standpoint, our findings suggest that many instructors see GenAI as a potentially powerful enhancement to classroom activities, beyond simply providing feedback or generating illustrative examples, and recognize its capacity to enrich students’ critical thinking, creativity, and engagement. For example, some instructors in our study envision GenAI-based assignments in which students compare AI-generated solutions with their own work to identify gaps or biases, thereby developing higher-order analytical skills. However, incorporating such practices can be complex: instructors might need to adjust traditional assessment criteria to account for AI-generated content, as well as ensure that students cultivate a discerning mindset when interacting with AI tools. Realizing GenAI’s pedagogical promise calls for a shift in teaching methods, from merely transmitting information to facilitating critical engagement with AI-generated resources, while maintaining rigorous standards for learning outcomes and fairness.

\subsection{Supporting Calibrated Trust in GenAI Among Instructors} 
Our study has illuminated the landscape of trust and distrust in GenAI among instructors in higher education. We found that trust in GenAI is not merely the absence of distrust but a multifaceted sentiment shaped by a variety of factors, including familiarity with technology, pedagogical alignment, and ethical considerations. Instructors in our study who exhibited high trust in GenAI appreciated its potential to enhance teaching effectiveness and student engagement, often using a cautious yet optimistic ``trust, but verify'' approach to integrate these tools into their pedagogical practices. And yet, distrust was manifested by concerns over GenAI's potential to undermine academic integrity, propagate biases, and exacerbate the digital divide. Our findings indicate the complexity of trust dynamics in educational technology and highlight the need for tailored strategies to cultivate a balanced understanding and application of GenAI in teaching and learning environments.

From a theoretical perspective, our results resonate with extended models of technology acceptance (e.g., the Technology Acceptance Model, TAM ~\cite{davis1989technology}), wherein perceived usefulness and perceived ease of use are key drivers of adoption. While instructors who perceive GenAI as beneficial (usefulness) and straightforward to integrate (ease of use) may display higher trust, concerns over ethical or pedagogical misalignment—akin to distrust—can dampen acceptance. These insights suggest that fostering trust in GenAI requires addressing not only usability and effectiveness but also the broader social, ethical, and disciplinary contexts within which teaching unfolds. At the same time, instructors play a critical role in helping students calibrate their trust in GenAI, guiding them to form appropriate levels of trust and distrust. The guidance not only supports students’ ability to engage with GenAI effectively, but also strengthens the interpersonal trust between instructors and students~\cite{niedlich2021comprehensive}, which is fundamental to enhancing educational development~\cite{corrigan2010trust, brion2015can}.

\subsubsection{Designing Platforms to Foster (Dis)Trust Calibration among Instructors}
Prior work has started looking into approaches for trust calibration in AI, recognizing the importance of supporting calibrated trust to help users engage with AI appropriately. Various interventions have been explored to foster trust calibration, such as displaying confidence scores that indicate the likelihood of a model's outputs being correct to end users~\cite{zhang2020effect}, and designing adaptive monitoring systems that visually prompt users to calibrate their trust when over-trust or under-trust is detected during human-AI collaboration~\cite{okamura2020adaptive}. Additionally, efforts in explainable AI aim to improve the trustworthiness and transparency of AI, enabling users to adjust their trust based on a clearer understanding of AI models' decision-making process~\cite{sun2024trustllm}. Collectively, these interventions seek to create a more balanced and informed trust relationship between humans and AI by supporting trust calibration.
However, work that focuses on tailoring trust calibration mechanisms, specifically for instructors in higher education settings, is sparse. Related interventions in education have been primarily designed to support the student population, aiming to help students shape an ``appropriate'' level of trust to avoid blind trust. Yet, leaving the trust calibration among instructors behind could lead to adverse outcomes~\cite{faranda2015effects}. For example, without proper calibration mechanisms, instructors might either over-rely on these technologies without critical assessment (due to blind trust), or underutilize these powerful tools due to unjustified skepticism.

We propose some design implications to cultivate calibrated trust among instructors. For example, it would be interesting to explore how interactive platforms can be designed to enable instructors to experiment with GenAI tools within their specific teaching contexts. For example, when framed within TAM, the ``perceived ease of use'' dimension could be enhanced through user-friendly interfaces and guided modules, while ``perceived usefulness'' might be boosted by showcasing data-driven improvements to student engagement and learning outcomes. Indeed, some existing platforms~\cite{magicschool_ai, eduaide_ai} suggest the potential of applying AI-driven insights in teaching by providing tailored resources and functionalities for educational contexts while primarily targeting schools instead of university-level instruction.
These platforms typically include simulation environments where instructors can observe the effects of GenAI integration on student engagement and learning outcomes in real-time. The idea is that the instructors can iteratively adjust and refine their use of GenAI, fostering a deeper understanding and trust in the technology. In this case, experimentation with GenAI tools could also be used to provide instructors with data analytics capabilities, and real-time insights into how GenAI affects teaching efficacy and student learning, allowing instructors to make data-driven decisions about integrating technology into their classrooms. Example interventions might include tracking the usage and outcomes of GenAI applications and correlating them with student performance to highlight effective practices and areas needing adjustment.

Furthermore, future design may explore features such as the ``tell your story'' function, inspired by our showcased use cases where instructors in our study reflect on their use of GenAI in practice. Storytelling in educational contexts has been shown to be effective in enhancing reflective teaching practices~\cite{hensel1992storytelling}. In the context of GenAI teaching, new platform design can draw upon learning theories such as constructivism (i.e., constructing one's own knowledge through experiences and social interaction, rather than passively receiving knowledge)~\cite{cobern1993constructivism} to facilitate educators in sharing their personal narratives. Such platforms could allow educators to share their personal narratives about their interactions with GenAI tools and provide both reflective insights and practical feedback on applying GenAI tools in education. 

\subsubsection{Experimental Studies and Evidence-Based Adjustments to Examine (Dis)Trust Dynamics}
To effectively build and assess faculty trust in GenAI, more research is needed to conduct empirical studies to understand the dynamics of trust development among instructors. For example, future work can explore how different forms of training, support, and GenAI tool implementation affect faculty perceptions and acceptance, in addition to the current focus on assessing the effectiveness of interventions on students~\cite{amoozadeh2024trust}. Research could involve controlled trials where faculty members are exposed to various scenarios of GenAI use, including but not limited to the context of higher education, with systematic variations in training intensity, support levels, and the complexity of GenAI tasks. Leveraging on prior work on measuring changes in trust levels before and after interventions in other fields~\cite{zhang2020effect, okamura2020adaptive}, work that focused on trust in GenAI in teaching and learning among instructors could provide additional insights into what factors may most and least significantly impact faculty trust, as well as to identify strategies to support trust calibration among faculty. Moreover, future work may investigate TAM-based metrics (e.g., perceived usefulness, perceived ease of use, intention to adopt) alongside trust measures to help disentangle how classic acceptance factors align or conflict with instructors’ underlying levels of trust and distrust in the context of GenAI.

\subsubsection{Institutional Support and Training Programs to Enhance GenAI literacy}
At the institutional level, providing support through tailored training programs is critical to build calibrated trust among instructors and enhancing GenAI literacy, as our results show that familiarity with GenAI concepts significantly correlates with trust. Indeed, institutions have been exploring how training programs can be specifically designed to enhance faculty understanding of GenAI. For example, more advanced universities may provide detailed guidelines and GenAI sandboxes for instructors to experiment with~\cite{gaiHarvard}, while others may offer only basic instructions and suggestions~\cite{AsuGai2}. Many institutions, however, are still grappling with the uncertainties surrounding the effective integration of these technologies into their curricula~\cite{luo2024critical}. This disparity is further amplified by differences in resources and subject focus among universities, with well-funded institutions or those with expertise in related fields offering more robust support and gaining a competitive edge.

A related challenge is to develop metrics and gather feedback that can guide the continuous improvement of training programs and GenAI tools. This process should involve not only quantitative measures, such as survey data and usage statistics, but also qualitative feedback from faculty about their experiences and concerns. Such a feedback loop will enable institutions to iteratively refine their strategies and better align GenAI adoption and integration with faculty needs and institutional goals. Considering the contextual factors unique to different institutions is vital. Factors such as institutional culture, faculty demographics, and prior exposure to technology can influence the effectiveness of GenAI integration strategies~\cite{andreadis2024mixed, bodani2023knowledge}. 

One pedagogical implication is that GenAI literacy training should extend beyond technical knowledge to include critical pedagogy, ethics, and reflective teaching practices. By offering a culturally tailored curriculum that includes practical skill-building and theoretical underpinnings of AI, institutions may better promote ``calibrated trust'' that allows faculty to confidently incorporate GenAI into their pedagogy while maintaining rigorous standards of academic integrity and quality. We believe that by focusing on evidence-based strategies to foster calibrated (dis)trust, educational institutions can create a more supportive ecosystem for faculty, which could help their confidence in using GenAI technologies while safeguarding academic integrity and pedagogical quality.

\subsubsection{Exploring the Nuances of ``Blind Distrust'': Context-Specific Trust and Distrust in GenAI}
Recall that some instructors in our study expressed a sense of strong distrust towards the use of GenAI, specifically in higher education contexts; it is worth reflecting that it might be that such a sense of distrust is situated in particular aspects,  rather than a blanket distrust of GenAI technology as a whole. 
Furthermore, while our proposed strategies to cultivate calibrated trust are aimed at those already somewhat open to GenAI, they might not adequately address the concerns of those with long-established distrustful or overly trusting views. These individuals may be less likely to engage with materials that challenge their preconceptions. Future research may consider focusing on these extremes of trust (and distrust). Understanding these nuanced attitudes toward GenAI could reveal the underlying factors driving \textit{blind} distrust that might be rooted in ideological disputes~\cite{apple2009routledge}, including power and control dynamics within educational and societal contexts~\cite{selwyn2013distrusting}.

\subsection{Ethical Considerations in GenAI-Enabled Higher Education}
Our results reveal that several instructors, such as P8 and P45, had some concerns about GenAI’s environmental impacts, especially its high energy and water usage. These findings align with broader research on the extensive computational resources needed to train and deploy advanced AI models~\cite{hoffmann2022training, mcdonald2022great}. Recent studies warn that GenAI’s growth could one day rival the energy consumption of entire nations~\cite{Crawford_2024}, prompting educators and administrators to reexamine whether the benefits of these technologies justify their substantial carbon footprints~\cite{crawford2021atlas}. The environmental impacts of GenAI suggest that equipping learners with a deeper understanding of GenAI’s resource demands may be critical, both inside and outside the classroom.

Furthermore, concerns over GenAI’s sustainability also intersect with social justice issues, as communities that already suffer environmental disadvantages stand to be disproportionately harmed~\cite{birhane2021algorithmic}. When institutions that prize eco-friendly initiatives rapidly scale up GenAI use, they may inadvertently perpetuate inequities in regions with fewer resources to support such infrastructure. Such tension becomes especially stark on campuses that simultaneously champion green agendas and celebrate AI-driven advancements. Allowing GenAI to flourish without acknowledging its ecological repercussions risks not only tarnishing institutional commitments to sustainability but also further alienating vulnerable groups.

To reconcile GenAI’s promise with these complex ethical challenges, higher education institutions might adopt multi-pronged strategies. For example, they may strengthen formal policies that explicitly link AI initiatives to sustainability goals, potentially partnering with tech providers that prioritize greener computing practices~\cite{strubell2020energy}. Second, educators can explore lower-resource AI solutions, such as model distillation~\cite{gou2021knowledge} and edge computing~\cite{cao2020overview}, which may reduce energy consumption without sacrificing performance~\cite{xu2024survey}. Finally, institutions may need to promote targeted training and awareness campaigns so that faculty and students can recognize how choices around AI usage, platform selection, model size, and frequency of queries translate into environmental impacts. By doing so, these measures may help the academic community work toward bridging the gap between technological innovation and equitable, environmentally mindful practices.

\section{Limitations}  
While our empirical study offers valuable insights into instructors' current practices, perceptions, and varying levels of trust and distrust toward GenAI in higher education, it comes with several limitations. 
First, the generalizability of our findings is constrained by the sample size, potentially limiting their applicability in different countries or educational contexts. Future research should include diverse stakeholders and educational settings, such as K-12 and other sociocultural contexts, to gain a more comprehensive understanding of GenAI’s impact on education. Furthermore, the cross-sectional design of our survey—capturing a single moment in time without longitudinal tracking—restricts our ability to ascertain how instructors’ attitudes and uses of GenAI may evolve. More studies are needed to investigate these changing dynamics and explore the long-term implications of GenAI in education.

\section{Conclusion}
Through a survey study with 178 participants from a single U.S. university, this study examines how trust and distrust in GenAI are manifested among instructors in higher education. Our findings reveal that while many instructors considered themselves familiar with GenAI, their actual direct experiences with and application of GenAI in educational tasks are limited. Moreover, we observed that trust and distrust among instructors, though related, were distinct concepts that can co-exist, offering a nuanced understanding of (dis)trust dynamics in GenAI. Thus, future research should closely examine the constructs of not only trust but also distrust in GenAI in order to fully understand their implications and interrelations. Moreover, our findings showcase the variety of factors that could contribute to trust and distrust formation around GenAI, though some cases indicate a sense of blind distrust. Many instructors in our study took ``trust, but verify'' approaches, allowing for the development of calibrated trust and distrust in both their own and their students' use of GenAI. Moving forward, it is essential to develop strategies and interventions that not only cultivate calibrated trust among educators but also address the root causes of distrust, promoting a balanced and ethical integration of GenAI into educational practices.

\section*{Statements on Open Data and Ethics}
Our Institutional Review Board approved this study with ID: PHSC-2023-12-01-16692. Informed consent was obtained from all participants. The anonymized survey questions, raw data, and data analysis code are publicly available on \href{https://osf.io/gve2s/?view_only=ad97f7079ef543d6894f68349ee03ae9}{OSF}.

\section*{Conflict of Interest}
The authors declare that they have no conflicts of interest.

\section*{Acronyms}

\begin{tabbing}
    GenAI \hspace{1cm} \= Generative Artificial Intelligence \\
    STEM \> Science, technology, engineering, and mathematics \\
    ITS \> Intelligent Tutoring System \\
    IRB \> Institutional Review Board \\
    TAM \> Technology Acceptance Model
\end{tabbing}

\begin{acks}
This work is supported by the National Science Foundation for support under award no. NSF-2418582. We thank the Provost's Office at William \& Mary, particularly Pamela L. Eddy, Associate Provost for Faculty Affairs and Development for survey design advice and dissemination effort, and Pablo Yanez, Senior Program Manager at the Studio for Teaching \& Learning in Innovation (STLI) at William \& Mary for his insightful survey feedback. We also appreciate our anonymous reviewers' constructive reviews.
\end{acks}

\bibliographystyle{ACM-Reference-Format}
\bibliography{main}

\appendix

\section{Codebook}
\label{appendix:codebook}

\subsection{Theme 1: GenAI Usage}
\textbf{Description:} This theme captures the surveyed instructors' current practices, experiences, and perspectives regarding the use of GenAI in their teaching in \autoref{tab:theme_1}.

\begin{table*}[h!]
\centering
\footnotesize
\begin{tabular}{@{}p{0.1\textwidth}p{0.3\textwidth}p{0.3\textwidth}p{0.25\textwidth}@{}}
\toprule
\textbf{Code} & \textbf{Description} & \textbf{Key Characteristics} & \textbf{Example Quotes} \\ 
\midrule
\textbf{GenAI Policy} & Refers to surveyed instructors’ current implementation of GenAI-related practices in their classes or their views on how policies governing GenAI use should be formulated. & Perspectives on integrating, restricting, or banning GenAI use, and clearly defined rules or practices concerning GenAI use in coursework. & \emph{``I forbid its use in my classes and therefore consider its use a violation of the honor code.''} (P79) \\ 
\midrule
\textbf{Use Case} & Refers to surveyed instructors’ current usage of GenAI for their own educational purposes or observed usage by students in academic settings. & Instructors’ or students’ use of GenAI in tasks like creating materials, solving problems, or enhancing learning experiences. & \emph{``I sometimes design homework/test problems using ChatGPT.''} (P12) \\ 
\midrule
\textbf{Negative Example} & Refers to instances where GenAI use is perceived as problematic, such as undermining academic integrity or leading to misuse. & Cases of cheating, plagiarism, or other unethical uses of GenAI in academic contexts. & \emph{``Students use ChatGPT to write essays, which frequently get bad grades anyway because they don't follow the prompt.''} (P23) \\ 
\bottomrule
\end{tabular}
\caption{Codes under Theme 1: GenAI Usage}
\label{tab:theme_1}
\end{table*}

\subsection{Theme 2: Impact of GenAI}
\textbf{Description:} This theme explores the surveyed instructors' concerns about how GenAI affects students and instructors in \autoref{tab:theme_2}.

\begin{table*}[h!]
\footnotesize
\centering
\begin{tabular}{@{}p{0.1\textwidth}p{0.3\textwidth}p{0.3\textwidth}p{0.25\textwidth}@{}}
\toprule
\textbf{Code} & \textbf{Description} & \textbf{Key Characteristics} & \textbf{Example Quotes} \\ 
\midrule
\textbf{Originality \& Creativity} & Refers to concerns that GenAI usage may diminish students' originality, individuality, and creative skill development in academic work. & Issues related to the loss of personal expression and reliance on GenAI lead to generic or homogenized outputs and hinder students’ ability to develop creative problem-solving and writing skills independently. & \emph{``The use of AI to write papers and complete assignments, which prevent the students from learning and improving their writing.''} (P29) \\ 
\midrule
\textbf{Integrity} & Refers to concerns about the ethical implications of GenAI usage, including its potential to facilitate academic dishonesty. & Issues related to cheating, plagiarism, or violations of academic integrity policies. & \emph{``I am concerned about students cheating in classes.''} (P65) \\ 
\midrule
\textbf{Over-Reliance} & Refers to concerns about students becoming overly dependent on GenAI, thereby missing opportunities for independent learning and skill development. & Issues related to students substituting personal effort and critical thinking with GenAI assistance. & \emph{``Students rely too heavily on it and do not take the time to critically evaluate the sources even though I teach them how to do this in class.''} (P31) \\ 
\midrule
\textbf{Mental Health} & Refers to the emotional and psychological challenges instructors or students experience in adapting to GenAI technologies, including stress, anxiety, and feelings of being overwhelmed. & Issues related to the pressure to adapt teaching practices, concerns about falling behind in technological understanding, and the need for support in navigating GenAI integration. & \emph{``I am not able to adapt quickly enough. I know my students are ahead of me on this. I would like to use Gen AI to set higher standards but it means a huge overhaul of how I teach. I need help because I don't even know what is possible.''} (P131) \\ 
\bottomrule
\end{tabular}
\caption{Codes under Theme 2: Impact of GenAI}
\label{tab:theme_2}
\end{table*}

\subsection{Theme 3: Perspectives on GenAI}
\textbf{Description:} This theme captures the surveyed instructors' perspectives on GenAI models, including their ethical, legal, and practical implications in \autoref{tab:theme_3}.

\begin{table*}[h!]
\footnotesize
\centering
\begin{tabular}{@{}p{0.08\textwidth}p{0.3\textwidth}p{0.3\textwidth}p{0.25\textwidth}@{}}
\toprule
\textbf{Code} & \textbf{Description} & \textbf{Key Characteristics} & \textbf{Example Quotes} \\ 
\midrule
\textbf{Copyright} & Refers to concerns regarding the ethical and legal implications of GenAI models being trained on copyrighted materials without proper permissions. & Issues related to intellectual property infringement, unauthorized use of scholarly or artistic works, and potential violations of copyright laws. & \emph{``Among other things, the models were built by infringing the intellectual property of scholars and artists including faculty on our own campus.''} (P19) \\ 
\midrule
\textbf{Misinfo} & Refers to concerns about GenAI producing and presenting inaccurate or false information, potentially misleading students and undermining their learning. & Issues related to the generation of misinformation, the authoritative tone of incorrect responses, and the impact on students' trust in reliable information sources. & \emph{``Gen AI provides false information in a confident tone and is dangerous for student learning.''} (P35) \\ 
\midrule
\textbf{Opposition} & Refers to strong resistance or rejection of the use of GenAI in academic settings, except in cases where it is the explicit subject of study. & Explicit disapproval of GenAI usage by students or faculty, concerns about its appropriateness in educational contexts, and calls for restrictions on its use. & \emph{``I am opposed entirely to the use of Gen AI, full stop, as a tool by students and faculty outside of contexts where Gen AI itself as a technological phenomenon is being studied.''} (P16) \\ 
\midrule
\textbf{Environment} & Refers to concerns about the environmental impact of GenAI, particularly its high energy consumption and contribution to carbon emissions. & Issues related to the ecological footprint of training and running GenAI models, including energy use, sustainability, and institutional responsibility. & \emph{``I also think as a campus we need to grapple with the massive environmental toll of Generative AI.''} (P20) \\ 
\midrule
\textbf{Human Rights} & Refers to concerns about the ethical implications of GenAI, including the exploitation of labor and other hidden costs associated with training these models. & Issues related to fair labor practices, ethical treatment of workers involved in the data labeling process, and raising awareness about the human costs behind GenAI technologies. & \emph{``Teaching everyone a better understanding of what these models actually can and cannot do, as well as of their hidden costs (... exploiting workers in the training process...), would be very valuable.''} (P8) \\ 
\bottomrule
\end{tabular}
\caption{Codes under Theme 3: Perspectives on GenAI}
\label{tab:theme_3}
\end{table*}

\end{document}